\documentclass[openacc]{rsproca_new}%%%%where rsproca is the template name
\usepackage{amsmath}
\usepackage{graphicx}
\usepackage[numbers,sort&compress]{natbib}
\graphicspath{{fig/}}

\titlehead{Review}
\begin{document}
    \title{Modern Time-Series and Spectral Methods for Analyzing Solar and Stellar Oscillatory Signals}
    \author{%%%% Author details
        Ding Yuan$^{1}$ and Song Feng$^{2}$}
    %%%%%%%%% Insert author address here
    \address{$^{1}$School of Aerospace, Harbin Institute of Technology, Shenzhen, Guangdong, China \\
        $^{2}$Faculty of Information Engineering and Automation, Kunming University of Science and Technology, Kunming, Yunnan, China
    }
    
    %%%% Subject entries to be placed here %%%%
    \subject{Signal Processing, Solar Physics, Stellar Physics}
    
    %%%% Keyword entries to be placed here %%%%
    
    %%%% Insert corresponding author and its email address}
\corres{Ding Yuan \\
    \email{yuanding@hit.edu.cn}\\
    Song Feng
    \email{feng.song@kust.edu.cn}
}

%%%% Abstract text to be placed here %%%%%%%%%%%%
\begin{abstract}
Time-series analysis plays a central role in understanding oscillatory and wave phenomena in solar and stellar atmospheres. However, astrophysical observations are inherently affected by instrumental noise, non-stationary dynamics, and uneven sampling. This review provides a comprehensive overview and comparative analysis of principal methods for detecting and characterizing periodicities in solar and stellar signals. We cover Fourier-transform-based transforms, nonlinear- fitting-based methods (Lomb--Scargle periodogram), time-frequency methods (wavelet and synchrosqueezed transforms), and adaptive decomposition techniques (Empirical Mode Decomposition). Advanced statistical significance tests, including false-alarm probability, autoregressive models, and Bayesian Markov Chain Monte Carlo (MCMC) approaches, are discussed their practical limitations and misuse risks. Through comparative analysis using synthetic benchmarks, we provide guidelines for selecting methods based on signal stationarity, sampling regularity, and noise characteristics. Finally, we outline future directions that integrate Bayesian inference with time-frequency analysis to achieve both statistical rigor and temporal localization in studying non-stationary solar and stellar oscillations.
\end{abstract}

\maketitle

\section{Introduction}

The study of oscillations and wave phenomena in the Sun and other stars has become a cornerstone of modern astrophysics, providing crucial diagnostics of plasma-magnetic field coupling, stellar interior dynamics, and energy transport mechanisms across multiple spatial and temporal scales \cite{khomenko2015oscillations, 2016SoPh..291.3143V, 2016LRSP...13....2B}. 
Observational signals---from helioseismic oscillations in the photosphere to magnetohydrodynamic (MHD) waves in the corona---encode rich physical information about local and global processes operating within magnetized plasma environments \cite{1983Natur.304..689G, 2009A&A...493..259I}. 
Accurate extraction and interpretation of such oscillatory features are therefore fundamental not only to solar and stellar physics but also to broader applications such as space-weather forecasting and stellar evolution modeling \cite{2016SoPh..291.3143V, 2018SSRv..214...45M}.

%\textbf{Most oscillation detection, characterization, and significance-testing methods in solar and stellar physics are fundamentally formulated for one-dimensional (1D) time series, typically extracted from individual pixels, slits, or spatially averaged regions in imaging and spectroscopic observations \cite{2012RSPTA.370.3193D}. These 1D formulations provide the methodological foundation for spatio-temporal diagnostics and are commonly applied to higher-dimensional datasets.}

 Classical time- and frequency-domain analysis techniques form the foundation of oscillation diagnostics. 
In the time domain, autocorrelation (ACF) and partial autocorrelation (PACF) functions have long served as intuitive tools for quantifying periodicity and temporal memory in solar and stellar time series \cite{2006MNRAS.369.1491R, 2010A&A...511A..46M}. 
In the frequency domain, Fourier transform \citep{cooley1965algorithm} and Lomb--Scargle periodogram (LSP) \citep{1976Ap&SS..39..447L, 1982ApJ...263..835S} techniques are extensively employed to detect stationary oscillations, such as the well-known five-minute solar oscillations and low-degree $p$-modes \cite{2017SPD....4810904M, 2018ApJ...855...65H}. 
Nevertheless, both ACF/PACF and traditional Fourier-transform based methods rely on the assumption of signal stationarity and therefore struggle to resolve transient or evolving oscillatory behavior.

The continuous wavelet transform (CWT) \cite{1998BAMS...79...61T} enables joint localization in time and frequency, making it highly effective for detecting transient oscillations in solar flares, coronal loops, and sunspot atmospheres \citep{2009Sci...323.1582J}. 
Empirical Mode Decomposition (EMD), combined with the Hilbert--Huang Transform (HHT), provides a fully data-driven means of decomposing nonlinear and non-stationary signals, revealing intrinsic oscillatory modes without assuming a fixed basis \cite{1998RSPSA.454..903H, wu2004study, 2011SoPh..269..439B}. 
More recently, the synchrosqueezed wavelet transform (SWT) \citep{daubechies2011synchrosqueezed, 2018ApJ...856L..16W} has been introduced to sharpen time-frequency representations, offering enhanced resolution for multi-component solar oscillations.

A central challenge in analyzing solar and stellar oscillatory signals lies in their contamination by instrumental noise, atmospheric disturbances, and stochastic astrophysical backgrounds, often coupled with pronounced non-stationarity and intermittency \cite{1976Ap&SS..39..447L, 1982ApJ...263..835S, 2005A&A...431..391V, 2010MNRAS.402..307V, 2016ApJ...825..110A, 2015ApJ...798..108I}. 
Advanced statistical frameworks have therefore been developed to account for noise and to assess the significance of oscillatory detections. 
False-alarm probability (FAP) methods \citep{1976Ap&SS..39..447L, 1982ApJ...263..835S} and autoregressive (AR) noise models \citep{1998BAMS...79...61T} are widely employed to characterize red-noise or power-law backgrounds in solar and stellar data. 
More recently, Bayesian inference combined with Markov Chain Monte Carlo (MCMC) sampling has enabled rigorous uncertainty quantification, posterior exploration, and model comparison in the frequency domain \cite{2015ApJ...798....1I, 2020Ap&SS.365...40L, 2023ApJ...944...16G}. 
These developments reflect a broader paradigm shift---integrating deterministic signal-processing techniques with probabilistic inference to achieve more robust and interpretable analyses of complex solar and stellar observations.

This review presents a comprehensive synthesis of contemporary methodologies for processing solar and stellar oscillatory signals. 
Section~\ref{sec:fundamentals} discusses the fundamental characteristics of these signals and observational sampling strategies, emphasizing their intrinsic nonstationarity and multi-periodic behavior. 
Sections~\ref{sec:fourier}--\ref{sec:emd} review key time- and frequency-domain methods, including the Fourier transform, nonlinear fitting methods, wavelet transform, and empirical mode decomposition. 
Since solar observations are strongly affected by stochastic backgrounds, Section~\ref{sec:noise} introduces three major approaches for noise estimation and significance testing. 
Section~\ref{sec:comparison} provides a systematic comparison of these methods using synthetic solar-like signals. 
Finally, \textbf{Section~\ref{sec:discussion} }discusses the relative merits of each approach and outlines future research directions.

\section{Fundamentals of Time Series and Spectral Analysis}
\label{sec:fundamentals}

\subsection{Sampling, Frequency Resolution, and the Nyquist Limit}

The analysis of solar and stellar oscillations critically depends on the temporal sampling of observational time series. 
For a uniformly sampled signal with a cadence (sampling interval) $\Delta t$ and a total duration $T = N \Delta t$, where $N$ is the total number of samples, the fundamental frequency resolution of any Fourier-transform based spectral analysis is
\begin{equation}
    \Delta f = \frac{1}{T}.
    \label{eq:frequency_resolution}
\end{equation}
A longer observing baseline $T$ therefore enhances the ability to resolve closely spaced oscillation modes, such as the low-degree $p$- and $g$-modes detected by long-duration missions like \textit{SOHO} and \textit{Kepler} \cite{1995SoPh..162....1D, 2010ApJ...713L..79K, 1983Natur.304..689G}. 
Conversely, short but high-cadence observations---typical in studies of solar flares, coronal-loop oscillations, and wave trains---provide better temporal localization at the cost of reduced frequency precision \cite{harvey1996global,2010A&A...511A..46M}. 
This reflects the fundamental trade-off between temporal and spectral resolution.

According to the Nyquist sampling theorem, the highest unambiguous frequency that can be detected in a discretely sampled signal is
\begin{equation}
    f_{\mathrm{Nyq}} = \frac{f_s}{2} = \frac{1}{2\Delta t},
    \label{eq:nyquist}
\end{equation}
where $f_s = 1/\Delta t$ is the sampling frequency. 
Spectral power at true frequencies above $f_{\mathrm{Nyq}}$ is \emph{aliased} and appears spuriously at lower frequencies, leading to false oscillatory peaks in the power spectrum. 
Aliasing effects are particularly severe in ground-based solar observations with data gaps or irregular sampling, where day--night cycles introduce systematic periodic artifacts. 
Common mitigation techniques include windowing, interpolation, or detrending before spectral estimation. 
In more complex cases, advanced approaches such as Lomb--Scargle periodograms or wavelet-based analyses can handle uneven sampling more robustly.

The Nyquist frequency defined in Eq.~(\ref{eq:nyquist}) strictly applies to uniformly sampled time series. For unevenly sampled data, such as those analyzed using the Lomb--Scargle periodogram, a unique Nyquist frequency is not rigorously defined. In practice, a ``pseudo-Nyquist'' frequency, often associated with the inverse of the median sampling interval, is sometimes adopted. Although Lomb--Scargle methods can formally explore frequencies beyond this average sampling limit, such analyses are more susceptible to aliasing and require careful interpretation.

Finally, the time-frequency uncertainty principle imposes an intrinsic limitation on all spectral analyses:
\begin{equation}
    \Delta t \, \Delta f \gtrsim \frac{1}{4\pi},
    \label{eq:uncertainty}
\end{equation}
which expresses the fundamental trade-off between temporal resolution and frequency precision: improving one necessarily degrades the other.

\subsection{Temporal Correlation}

To characterize periodicity and persistence in solar and stellar oscillations, time-domain diagnostics such as the autocorrelation function (ACF) and partial autocorrelation function (PACF) are widely employed. 
The ACF measures the linear correlation of a signal with its lagged copies, while the PACF isolates the direct correlation between $x_t$ and $x_{t-k}$ after removing intermediate dependencies. 
The normalized ACF is expressed as
\begin{equation}
    \rho(k) = \frac{\mathbb{E}[(x_t - \mu)(x_{t-k} - \mu)]}{\sigma^2},
    \label{eq:acf}
\end{equation}
where $\mu$ and $\sigma^2$ are the mean and variance of the signal, and $k$ is the time lag. 
For a purely random white-noise process, $\rho(k)$ decays to zero rapidly for $k > 0$, providing a baseline for identifying significant quasi-periodic behavior above random fluctuations. 
Hence, ACF analysis can effectively suppress uncorrelated white noise.

However, in solar and stellar contexts where the background often follows a red-noise or power-law spectrum, the ACF decays slowly, sometimes mimicking long-term correlations unrelated to true oscillations. 
The PACF, though less sensitive to indirect correlations, may also yield misleading results in such cases. 
Therefore, while ACF/PACF analysis remains a valuable first step for detecting temporal coherence, it must be complemented by frequency- or time-frequency-domain methods to distinguish physical periodicities from stochastic persistence.

\subsection{Noise Spectra of Solar and Stellar Signals}

Solar and stellar oscillatory signals are inherently non-stationary, exhibiting time-dependent amplitudes, frequencies, and variances \cite{2016ApJ...825..110A,2016LRSP...13....2B}. 
They are typically embedded in a turbulent background driven by magnetoconvection and MHD processes, which produces broad-band stochastic fluctuations. 
The power spectral density (PSD) of such noise commonly follows a power-law form,
\begin{equation}
    P(f) \propto f^{-\beta},
    \label{eq:power_law}
\end{equation}
where $\beta \approx 0$ corresponds to white noise and $\beta \gtrsim 1$ indicates red or power-law noise. 
This red-noise background enhances low-frequency power, often obscuring weak periodic components and complicating their detection through standard Fourier or wavelet analyses.

Physically, red-noise behavior reflects the long-memory correlations characteristic of solar convection and turbulent plasma flows. 
In the time domain, this manifests as a slowly decaying autocorrelation function, whereas in the frequency domain, it produces steep spectral slopes. 
Accurate modeling of this background---using autoregressive (AR), power-law, or Bayesian frameworks---is therefore essential to avoid false detections and to assign statistically meaningful confidence levels to oscillatory features.

%\subsection{From 1D Time Series to Spatio-Temporal Oscillation Diagnostics}
%\textbf{Although the methods reviewed in this paper are formulated for one-dimensional time series, they are widely used as building blocks for analyzing multi-dimensional solar and stellar observations. In imaging and spectroscopic data cubes (e.g., SDO/AIA, Hinode/EIS, IRIS), oscillatory analysis is typically performed by extracting 1D time series at each pixel, along a slit, or within spatially defined regions.}

%\textbf{Common extensions include pixel-by-pixel Fourier or wavelet mapping, construction of $k$--$\omega$ (wavenumber--frequency) diagrams, and spatial maps of oscillatory power, phase, and coherence. Such approaches enable the identification of propagating waves, standing modes, and spatially coherent oscillatory structures across the solar atmosphere. Comprehensive overviews of these spatio-temporal diagnostics and their physical interpretation are provided by \cite{2012RSPTA.370.3193D}.}

\section{Fourier-transform based Methods}
\label{sec:fourier}

Fourier-transform based analysis provides the mathematical foundation for nearly all modern signal-processing techniques used in solar and stellar physics. 
The fundamental idea is to decompose a time series into sinusoidal components, thereby expressing it as a sum of oscillations with distinct frequencies and amplitudes. 
This decomposition yields a global representation of how signal power is distributed across frequencies. 
A key assumption of all Fourier-transform based methods is \textit{stationarity}---that the statistical properties and frequency content of the signal remain constant over time. 
While this assumption holds approximately for global oscillations such as solar $p$-modes, it is often violated in transient or evolving phenomena.

\begin{figure}[ht!]
\centering
\includegraphics[width=0.95\textwidth]{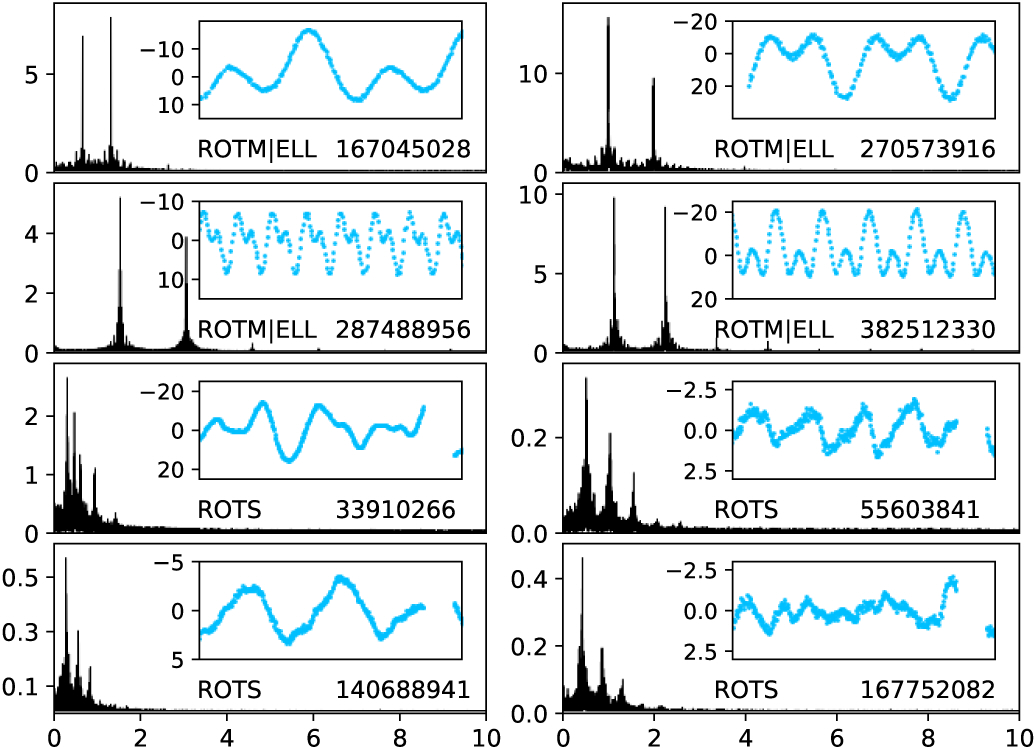}
\caption{Examples of rotationally variable stars. The horizontal axis represents frequency (in cycles per day, c/d), while the vertical axis shows brightness amplitude (in millimagnitudes, mmag). The TIC numbers in each panel correspond to the TESS Input Catalog identifiers. The main panels display the frequency spectra, and the insets illustrate the light curves over 5~days (upper four panels) and 10~days (bottom panel). As seen from these examples, rotational modulation typically manifests as one or several dominant peaks and their harmonics in the frequency domain, accompanied by relatively stable time-domain variability.
% Image reproduced with permission from \citep{2024A&A...688A..25S}.
 Image reproduced from \citep{2024A&A...688A..25S} under the Creative Commons Attribution 4.0 License (CC BY 4.0).} 
 \label{fig:fft}
\end{figure}

\subsection{Continuous and Discrete Fourier Transforms}

The Continuous Fourier Transform (CFT) provides the theoretical basis for frequency analysis by mapping a continuous signal $x(t)$ into its spectral domain:
\begin{equation}
    X(f) = \mathcal{F}\{x\} = \int_{-\infty}^{\infty} x(t)\, e^{-i 2\pi f t}\, dt,
    \label{eq:cft}
\end{equation}
where $X(f)$ represents the complex amplitude at frequency $f$. 
The inverse transform reconstructs the original signal as
\begin{equation}
    x(t) = \mathcal{F}^{-1}\{X\} = \int_{-\infty}^{\infty} X(f)\, e^{i 2\pi f t}\, df.
    \label{eq:icft}
\end{equation}
Although the CFT offers ideal spectral resolution, it assumes continuous sampling over infinite duration and thus cannot be directly implemented in digital analysis. 
Nevertheless, it provides the conceptual foundation for the discrete versions used in practice.

For uniformly sampled data $x_n$ ($n = 0, 1, \dots, N-1$) with sampling interval $\Delta t$, the Discrete Fourier Transform (DFT) is defined as
\begin{equation}
    X_k = \sum_{n=0}^{N-1} x_n\, e^{-i 2\pi k n / N}, \quad k = 0, 1, \dots, N-1,
    \label{eq:dft}
\end{equation}
where each frequency bin is
\begin{equation}
    f_k = \frac{k}{N \Delta t}, \quad \text{and} \quad \Delta f = \frac{1}{T} = \frac{1}{N \Delta t}.
    \label{eq:freq_bins}
\end{equation}
The DFT provides a finite, periodic approximation of the continuous spectrum and forms the computational basis of most spectral analyses in helio- and asteroseismology.

\subsection{Fast Fourier Transform and Periodogram Estimation}

The Fast Fourier Transform (FFT) \citep{cooley1965algorithm} is an efficient numerical algorithm for computing the DFT, reducing computational complexity from $O(N^2)$ to $O(N \log N)$. 
This efficiency enables high-resolution spectral estimation for large solar and stellar datasets.
In solar and stellar applications, FFT-based periodograms are widely used to detect periodicities associated with flare quasi-periodic pulsations (QPPs), coronal-loop oscillations, and stellar rotation signatures \cite{2015ApJ...798....1I}. 
A representative example is shown in Figure~\ref{fig:fft}, where the brightness variations of rotationally variable stars observed by \textit{TESS} are decomposed into frequency components using FFT analysis. 
The distinct spectral peaks and their harmonics correspond to the stellar rotation frequency and its multiples, demonstrating how Fourier methods transform time-domain variability into interpretable frequency-domain signatures.

The Fourier power spectrum or periodogram is defined as
\begin{equation}
    P(f_k) = \frac{1}{N} \left| \sum_{n=0}^{N-1} x_n\, e^{-i 2\pi f_k n \Delta t} \right|^2,
    \label{eq:periodogram}
\end{equation}
which represents the distribution of oscillatory power across frequencies. 
The reliance on stationarity and uniform sampling can lead to spectral artifacts when applied to irregular or transient data.
Despite its simplicity, the periodogram is a biased and high-variance estimator---particularly for short, noisy, or non-stationary signals.

\subsection{Windowing and Spectral Leakage}
\label{sec:windowing}

In practice, observational time series are finite and often truncated, introducing discontinuities at the boundaries. 
This truncation causes \textit{spectral leakage}, where power from one frequency spreads into adjacent frequencies, distorting the true spectrum. 
To mitigate this, window functions are applied in the time domain before computing the FFT.

If $w_n$ represents a window function, the windowed DFT is
\begin{equation}
    X_k^{(w)} = \sum_{n=0}^{N-1} w_n\, x_n\, e^{-i 2\pi k n / N}.
    \label{eq:windowed_dft}
\end{equation}
The simplest case, the rectangular (boxcar) window, assumes no tapering---it preserves amplitude but produces the strongest spectral leakage. 
In contrast, tapered windows such as the Hann or Hamming functions gradually reduce amplitudes near the signal edges, suppressing sidelobes and improving spectral purity at the expense of slightly broadening the main lobe. 
This trade-off between resolution and leakage is crucial for accurate spectral estimation in solar physics, where oscillatory signals often coexist with strong broadband noise.
Applying the window reduces spectral leakage and improves frequency localization, though at the cost of a slightly lower effective resolution. 
Such preprocessing is essential when analyzing finite or noisy datasets from instruments such as SOHO, SDO, or Kepler.

Fourier-transform based spectral analysis remains a cornerstone of helio- and asteroseismic research due to its theoretical rigor and computational efficiency. 
It is ideally suited to stationary or quasi-stationary signals with well-defined periodicities. 
However, it faces challenges in non-stationary or unevenly sampled data, where spectral leakage, aliasing, and noise sensitivity limit interpretability.

\section{Nonlinear-Fitting based Methods}
\label{sec:nonlinear}

Traditional spectral analysis techniques such as the Fourier transform assume uniformly sampled and stationary signals. 
However, solar and stellar observations often suffer from irregular sampling due to instrumental gaps, day--night cycles, or observational constraints. 
To address these challenges, two frequency-domain methods based on nonlinear or least-squares fitting have been developed: the date-compensated Fourier transform (DCFT) \citep{1981AJ.....86..619F} and the Lomb--Scargle periodogram (LSP) \citep{1976Ap&SS..39..447L,1982ApJ...263..835S}. 
These techniques generalize the Fourier framework to unevenly sampled data and enable robust estimation of periodicities without resampling or interpolation.

\subsection{Date-compensated Fourier Transform (DCFT)}

The date-compensated Fourier transform (DCFT) \cite{1981AJ.....86..619F} was one of the earliest attempts to generalize the Fourier transform to unevenly spaced data. 
It compensates explicitly for the time irregularity (``date compensation'') by modifying the standard Fourier coefficients to account for nonuniform sampling. 
For an observation set $\{x_i, t_i\}$, the DCFT computes the amplitude at frequency $\omega$ using:
\begin{equation}
    A(\omega) = \frac{\sum_i x_i \cos \omega t_i}{\sum_i \cos^2 \omega t_i}, 
    \qquad
    B(\omega) = \frac{\sum_i x_i \sin \omega t_i}{\sum_i \sin^2 \omega t_i},
    \label{eq:dcft_coeff}
\end{equation}
and defines the power spectrum as
\begin{equation}
    P_{\mathrm{DCFT}}(\omega) = \frac{1}{2}(A^2(\omega) + B^2(\omega)).
    \label{eq:dcft_power}
\end{equation}
The ``date compensation'' arises from the normalization terms in the denominators, which correct for uneven temporal sampling and prevent bias in amplitude estimation.
DCFT is conceptually simpler but less statistically rigorous, as it lacks a clear false-alarm probability framework. 
Nevertheless, it provides a computationally efficient approach for exploratory frequency analysis in irregularly sampled solar datasets, especially when only approximate detection of dominant periods is required. 
It has been historically applied in early helioseismology and stellar activity studies, particularly before the development of the modern LSP framework.

\subsection{Lomb--Scargle Periodogram (LSP)}

The LSP \cite{1976Ap&SS..39..447L,1982ApJ...263..835S} is a least-squares spectral estimator designed for unevenly spaced observations. 
Unlike the standard discrete Fourier transform, which directly applies sinusoidal basis functions to uniformly sampled data, the LSP determines the sinusoidal coefficients that best fit the observed data in a least-squares sense. 
For an unevenly sampled signal $\{x_i, t_i\}$, the normalized periodogram is defined as:
\begin{equation}
    P_{\mathrm{LS}}(\omega) = \frac{1}{2\sigma^2}
    \left[
    \frac{\left[\sum_i (x_i - \bar{x}) \cos \omega (t_i - \tau)\right]^2}{\sum_i \cos^2 \omega (t_i - \tau)}
    +
    \frac{\left[\sum_i (x_i - \bar{x}) \sin \omega (t_i - \tau)\right]^2}{\sum_i \sin^2 \omega (t_i - \tau)}
    \right],
    \label{eq:lsp}
\end{equation}
where $\bar{x}$ and $\sigma^2$ are the mean and variance of the data, $\omega$ is the angular frequency, and $\tau$ is a phase offset defined to ensure orthogonality between sine and cosine terms.
The LSP provides an unbiased and statistically rigorous way to identify periodicities in irregularly sampled time series. 
In addition, extensions such as the generalized LSP incorporate weights, trends, or multiple harmonics, improving robustness against noise and long-term variability \cite{2009A&A...496..577Z}.

\begin{figure}[ht!]
    \centering
    \includegraphics[width=0.95\textwidth]{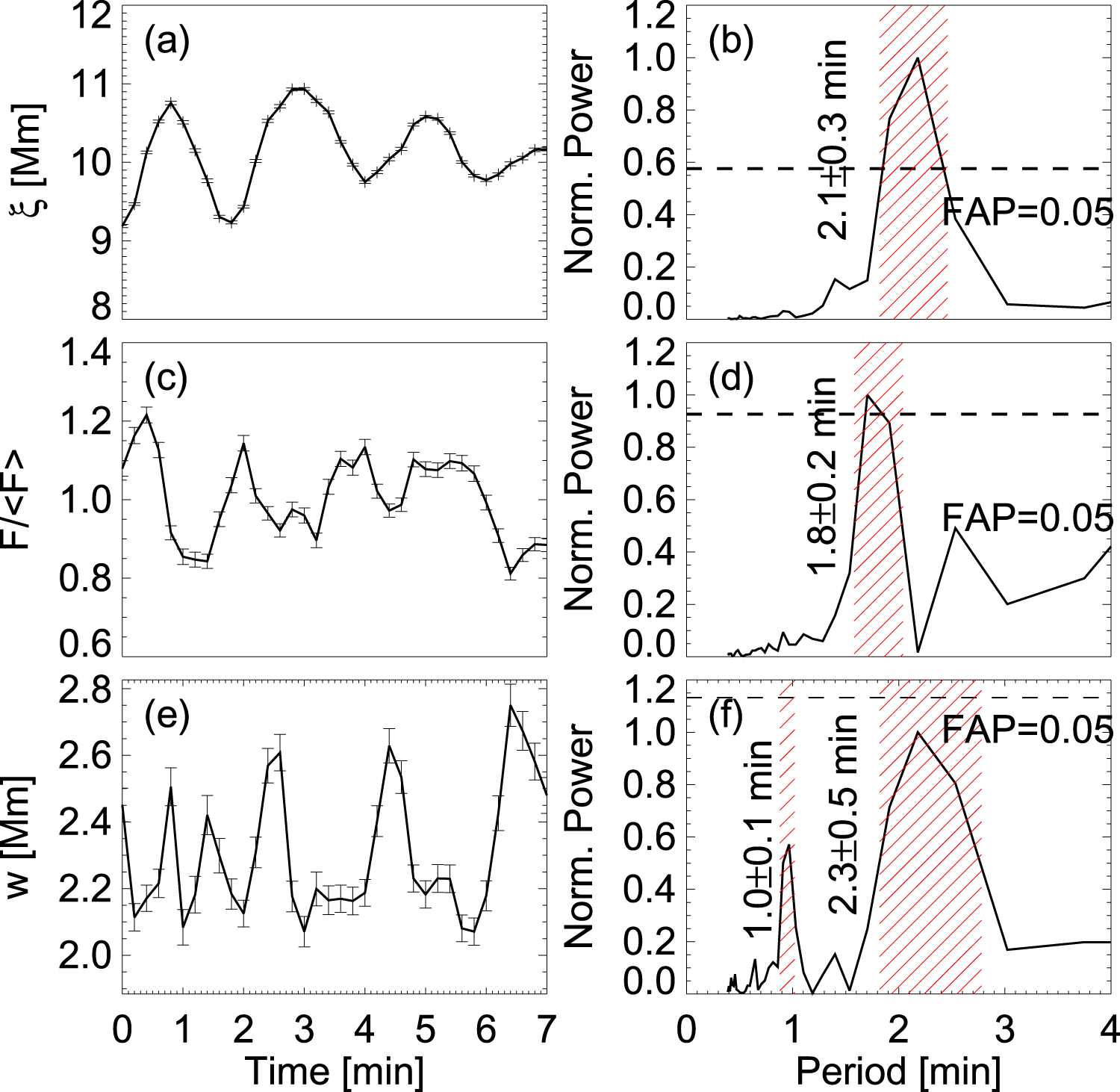}
\caption{Illustration of the Lomb--Scargle periodogram (LSP) analysis applied to solar coronal loop oscillations. Panels~(a), (c), and (e) show the time series of loop displacement~($\xi$),   normalized emission flux~($F / \langle F \rangle$), 
        and loop width~($w$), respectively, measured at $s = 0.5L_0$ in Event~V. 
        Panels~(b), (d), and (f) display the corresponding Lomb--Scargle power spectra. 
        The dashed lines indicate the 0.05 false alarm probability (FAP) level, 
        while the hatched regions highlight the significant oscillation periods identified in each signal. 
        This figure exemplifies how the LSP method can detect periodic signatures 
        in unevenly sampled or noisy time series data. 
        % (Modified from \cite{2016ApJS..223...24Y})
         Image reproduced from  \citep{2016ApJS..223...24Y}. \textcopyright AAS. Reproduced with permission.
    }
    \label{fig:lsp}
\end{figure}

The LSP is a widely used technique for detecting and characterizing periodicities in unevenly sampled or noisy astronomical time series. 
Unlike the classical Fourier transform, which assumes uniform sampling, the LSP provides an unbiased estimate of the spectral power even when data gaps are present. 
Figure~\ref{fig:lsp}, modified from \cite{2016ApJS..223...24Y}, illustrates a typical application of the LSP method to coronal loop oscillations. 
The left panels show time series of loop displacement, normalized intensity, and loop width, while the right panels present their corresponding LSP spectra. 
Prominent peaks in the periodograms, highlighted by hatched areas, indicate statistically significant oscillation periods, with the dashed lines marking the 0.05 false alarm probability threshold. 
This example demonstrates how the LSP effectively extracts oscillatory signals from complex solar observational data and quantitatively assesses their significance. 
In the context of solar and stellar oscillation studies, such LSP analysis plays a crucial role in identifying dominant frequencies, mode structures, and temporal evolution of oscillations when direct Fourier analysis is limited by irregular cadence or instrumental interruptions.

The primary advantages of LSP are its ability to handle irregular sampling without interpolation, its direct statistical interpretation (false-alarm probabilities), and its connection to least-squares sinusoidal fitting. 
However, its sensitivity to correlated (red) noise and the assumption of sinusoidal oscillation shapes can limit its effectiveness for strongly nonlinear or transient solar signals. 
For such cases, time-frequency or adaptive decomposition methods are often more appropriate.

Both LSP and DCFT can be interpreted as nonlinear fitting or generalized least-squares spectral estimation methods. 
They differ from traditional FFT-based approaches in that they fit sinusoidal models directly to the observed data rather than projecting the signal onto uniformly spaced frequency bins. 
This makes them more flexible in dealing with irregularly sampled or gapped datasets---a common situation in solar and stellar observations.

However, neither method is inherently adaptive to time-varying frequencies. 
They provide global frequency spectra and are thus most effective for stationary or quasi-stationary periodic signals. 
For signals exhibiting strong nonstationarity, such as flare oscillations or evolving coronal waves, wavelet or synchrosqueezed methods offer superior temporal localization.

\section{Wavelet-based Methods}
\label{sec:wavelet}

While Fourier-based techniques provide a global view of frequency content, they assume signal stationarity and therefore cannot effectively characterize transient or time-varying oscillations. 
In contrast, wavelet-based methods offer joint time-frequency localization, enabling the identification of transient oscillations, frequency drifts, and mode coupling processes commonly present in solar and stellar signals. 
Wavelet transforms can be categorized into two main forms: the discrete wavelet transform (DWT) and the continuous wavelet transform (CWT). 
Although both rely on the same underlying concept of representing a signal as the sum of localized wavelets, their mathematical properties, computational implementation, and practical applications differ significantly.
The DWT and CWT are complementary rather than competing tools. 
The DWT provides an efficient, non-redundant representation suitable for feature extraction, noise reduction, and hierarchical decomposition of stationary or quasi-stationary components. 
The CWT, on the other hand, provides a high-resolution, redundant time-frequency map ideal for detecting and characterizing non-stationary or transient oscillations in solar and stellar observations. 
However, it requires more computation and produces redundant information, since scales and translations are not discretized. 

In addition, the CWT suffers from edge effects---regions of reduced confidence near the beginning and end of the time series (known as the ``cone of influence'')---which defines the region of the wavelet power spectrum where edge effects become significant due to the finite length of the time series. Power within the COI is therefore considered unreliable and should be excluded from physical interpretation.

In practical applications, the DWT is often used for preprocessing---such as detrending or denoising---before applying the CWT for detailed spectral analysis.

\subsection{Continuous and Discrete Wavelet Transform (CWT/DWT)}

The continuous wavelet transform extends this framework to continuous scales and translations, providing a highly detailed time-frequency representation. 
For a signal $x(t)$, the CWT is defined as
\begin{equation}
    W(a,b) = \frac{1}{\sqrt{a}}\int_{-\infty}^{\infty} x(t)\psi^*\left(\frac{t-b}{a}\right)dt,
    \label{eq:cwt}
\end{equation}
where $\psi(t)$ is the mother wavelet, $a$ is the scale parameter (inversely related to frequency), and $b$ is the translation parameter representing time localization. 
The resulting wavelet power spectrum, $|W(a,b)|^2$, provides information on how oscillatory power evolves with both time and frequency.

The CWT method bridges the gap between time-domain and frequency-domain analyses. The key advantage of wavelet analysis lies in its ability to capture transient, non-stationary phenomena, which are ubiquitous in the solar atmosphere. 
They provide both temporal and spectral information, making them particularly valuable for studying the dynamic solar atmosphere where multiple oscillatory processes interact across a wide range of spatial and temporal scales.

In wavelet analysis, the Morlet mother wavelet has been widely adopted because of its close connection to Fourier analysis, allowing for an intuitive interpretation of frequency and phase. 
Using wavelet analysis, \cite{2010SoPh..262..373B} analyzed oscillations in solar magnetic loops observed with EUV imaging. \cite{2021ApJ...921..179L} employed wavelet diagnostics to identify quasi-periodic pulsations (QPPs) during the impulsive phase of a powerful flare. More recently, \cite{2009Sci...323.1582J} combined high-cadence imaging spectroscopy with wavelet analysis to detect Alfvén waves in the chromosphere, and \cite{2004A&A...418..313U} applied the method to coronal point oscillations observed by SOHO/CDS, EIT, and MDI. These examples demonstrate the broad applicability of wavelets to both transient flare-related oscillations and persistent coronal or chromospheric modes.

\begin{figure}[ht!]
    \centering
    \includegraphics[width=0.95\textwidth]{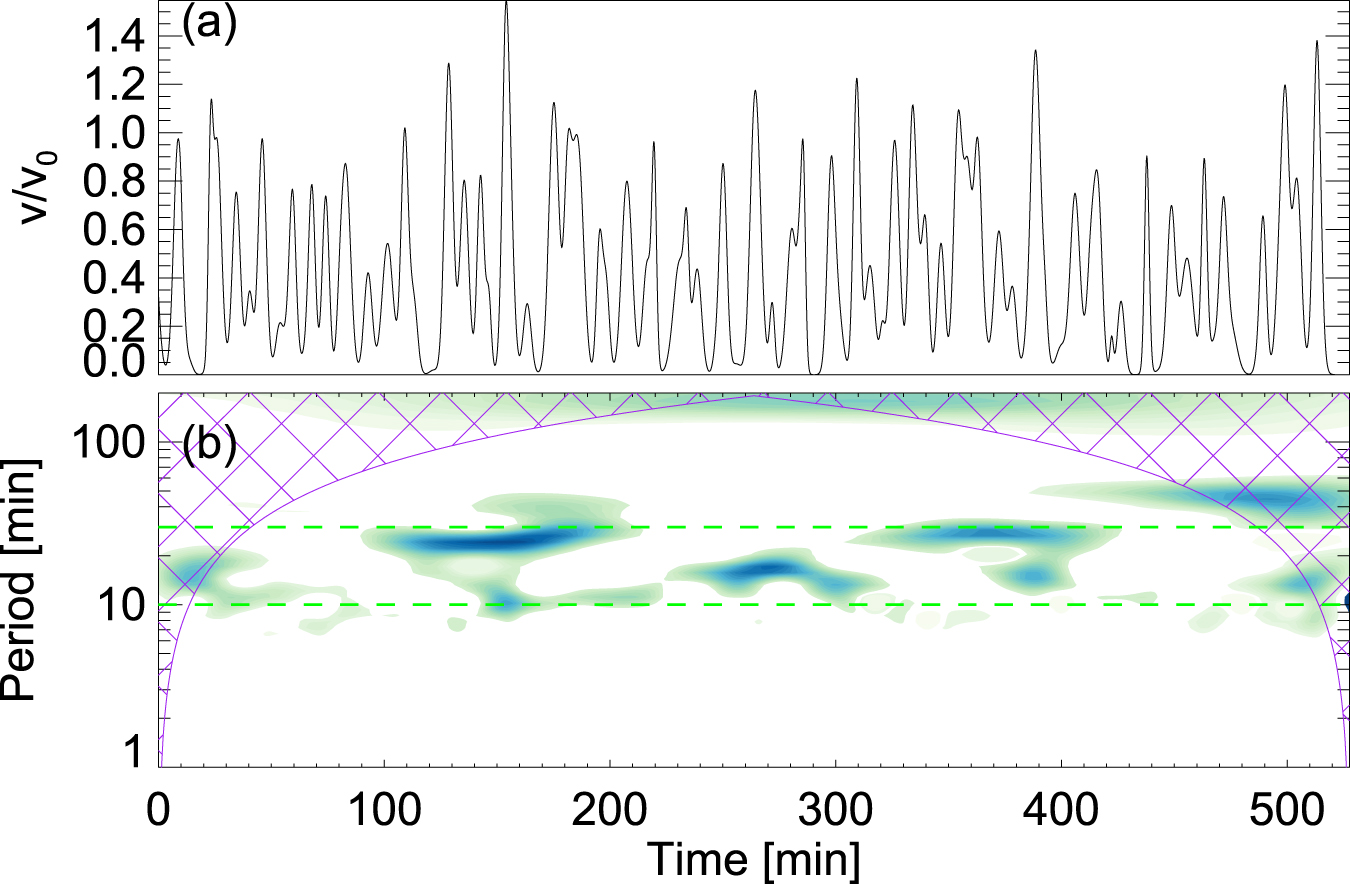}
    \caption{
        Demonstration of the continuous wavelet transform (CWT) applied to a stochastic finite-lifetime transient model. 
        Panel~(a) shows the synthetic time series, while panel~(b) presents the corresponding wavelet power spectrum. 
        The wavelet spectrum reveals quasi-periodic oscillations in the 10--30~minute range, enclosed by the green dashed lines. 
        The cross-hatched region marks the cone of influence, within which the power values become unreliable due to edge effects. 
        The model parameters are $P_0 = 300$~s and $\sigma_0 = 100$~s. 
        This figure illustrates how the CWT provides a localized time-frequency representation of non-stationary oscillations.
        % (Modified from \cite{2016ApJS..224...30Y})
         Image reproduced from \citep{2016ApJS..224...30Y}. \textcopyright AAS. Reproduced with permission.
    }
    \label{fig:cwt}
\end{figure}

The CWT is a powerful tool for analyzing non-stationary and transient oscillations in solar and stellar signals. 
Unlike the Fourier transform, which represents the signal purely in the frequency domain, the CWT retains both temporal and frequency localization, allowing researchers to track how oscillatory power evolves over time. 
Figure~\ref{fig:cwt}, modified from \cite{2016ApJS..223...24Y}, illustrates the application of the CWT to a synthetic stochastic finite-lifetime transient model. 
The time series in panel~(a) contains intermittent oscillations, and the corresponding wavelet power spectrum in panel~(b) reveals quasi-periodic components concentrated between 10 and 30~minutes, highlighted by green dashed boundaries. 
The cross-hatched region denotes the cone of influence, where edge effects render the spectral power unreliable. 
This example demonstrates how the CWT effectively captures both the period and temporal evolution of oscillatory signals, making it particularly suitable for studying solar coronal waves, transient brightenings, and other non-stationary astrophysical phenomena.

The discrete wavelet transform is based on the dyadic scaling and translation of a mother wavelet function $\psi(t)$, defined as
\begin{equation}
    \psi_{j,k}(t) = 2^{-j/2}\psi(2^{-j}t - k),
    \label{eq:dwt_basis}
\end{equation}
where $j$ and $k$ are integer indices representing scale (related to frequency) and translation (related to time), respectively. 
The DWT decomposes a discrete time series $x(t)$ into approximation and detail coefficients using iterative filtering operations:
\begin{equation}
    x(t) = \sum_{j}\sum_{k} c_{j,k}\psi_{j,k}(t),
    \label{eq:dwt_decomposition}
\end{equation}
where $c_{j,k}$ are the wavelet coefficients. 
Practically, this is implemented using a filter bank composed of low-pass and high-pass filters, recursively applied to yield multi-resolution decompositions.

The key advantages of DWT are its computational efficiency and compact representation of signal information. 
It provides an efficient framework for denoising, data compression, and feature extraction, making it valuable for processing large solar datasets such as high-cadence magnetograms or irradiance time series. 
However, the DWT only analyzes frequencies on a discrete dyadic grid and thus provides limited frequency resolution, making it less suitable for precise spectral characterization of oscillations.

\subsection{Synchrosqueezed Wavelet Transform}
\label{sec:swt}

The SWT is an enhanced time-frequency analysis method that builds upon CWT by introducing a frequency reallocation, or ``synchrosqueezing'' procedure. This process sharpens the spectral representation by concentrating energy that is otherwise dispersed across scales into their true oscillatory frequencies.

Given the wavelet coefficients $W_x(a,b)$ of a signal $x(t)$ at scale $a$ and time $b$, the local instantaneous frequency is estimated from the derivative of the phase of the wavelet transform:
\begin{equation}
    \omega_x(a,b) = -j \frac{\partial_b W_x(a,b)}{W_x(a,b)}.
    \label{eq:instantaneous_freq}
\end{equation}
Instead of interpreting the energy $|W_x(a,b)|^2$ directly in the time-scale plane, the synchrosqueezing step reallocates this energy into the time-frequency plane according to $\omega_x(a,b)$. The synchrosqueezed transform is then defined as
\begin{equation}
    T_x(\omega,b) = \int_{0}^{\infty} W_x(a,b) \, \delta\!\big(\omega - \omega_x(a,b)\big) \, \frac{da}{a^{3/2}},
    \label{eq:swt}
\end{equation}
where $\delta(\cdot)$ is the Dirac distribution. In practice, this integral is approximated by binning coefficients into discrete frequency intervals. The result is a sharpened time-frequency representation in which oscillatory components are concentrated into narrow frequency ridges rather than smeared across scales.

\begin{figure}[ht!]
    \centering
    \includegraphics[width=0.95\textwidth]{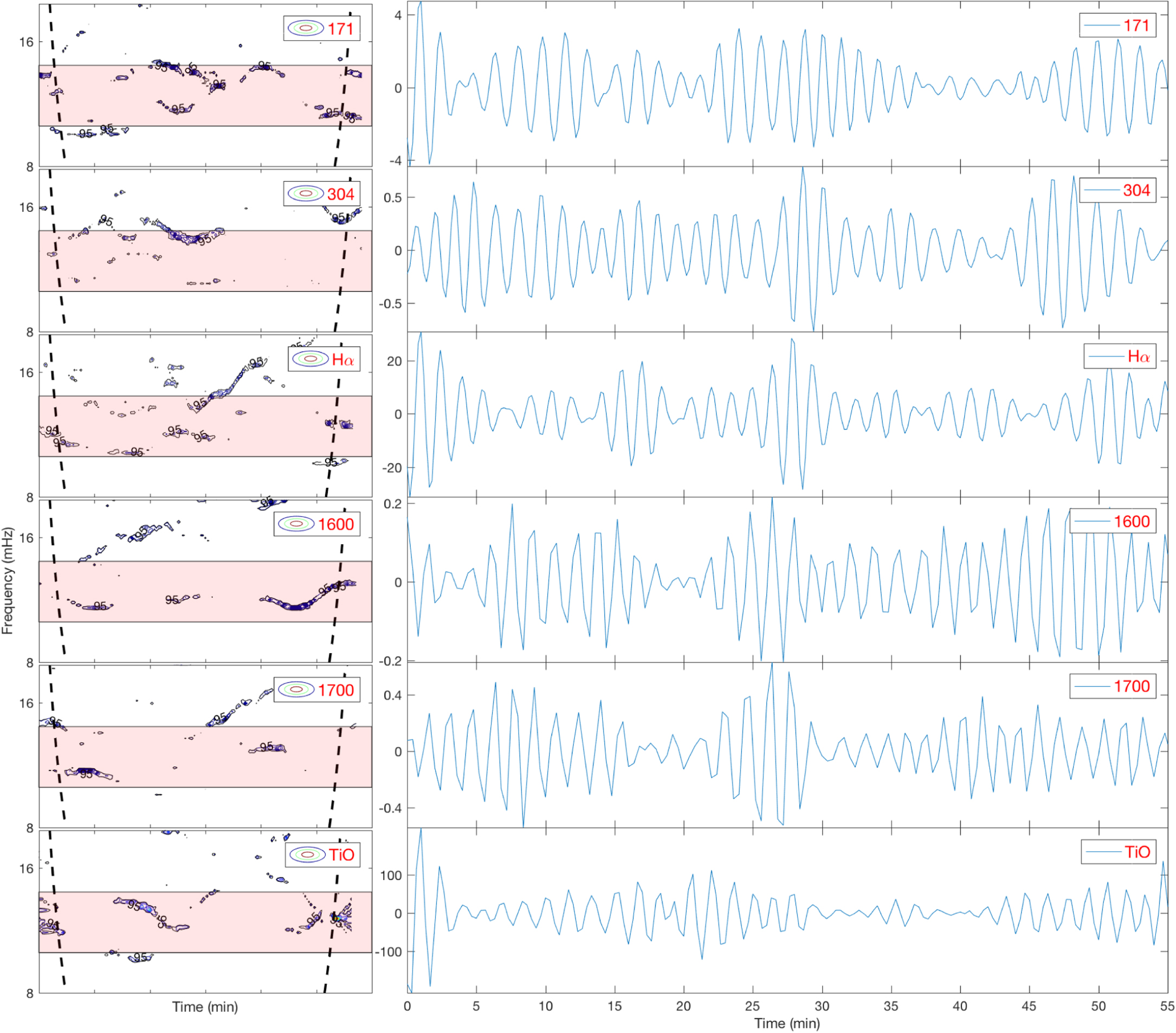}
    \caption{
        Left column: Synchrosqueezed Wavelet Transform (SWT) power spectra for different wavelength channels (171~\AA, 304~\AA, H$\alpha$, 1600~\AA, 1700~\AA, and TiO). 
        The dashed black curves indicate the cone of influence (COI) where edge effects become significant. 
        Right column: corresponding reconstructed temporal components obtained from inverse SWT, revealing coherent oscillatory behavior at multiple atmospheric heights. 
        % Image reproduced with permission from \citep{2018ApJ...856L..16W}.
        Image reproduced from \citep{2018ApJ...856L..16W}. \textcopyright AAS. Reproduced with permission.
    }
    \label{fig:swt}
\end{figure}

The SWT complements traditional CWT analysis by providing sharpened spectral localization. It is emerging as a valuable tool in solar and stellar time-frequency studies, particularly for disentangling multi-component oscillations and identifying weak periodicities in noisy astrophysical data. For instance, \cite{2020ApJ...902...64X} applied the SWT to study solar rotational periods, demonstrating a secular downtrend in the cycle-averaged rotation rate. \cite{2020MNRAS.494.4930D} employed the SWT to investigate quasi-biennial oscillations in the solar cycle, showing that the method can effectively isolate weak periodicities buried in background variability. \cite{2018ApJ...856L..16W} used the SWT on high-resolution observations from the New Vacuum Solar Telescope and SDO/AIA to identify minute-scale oscillations above a sunspot umbra. Figure \ref{fig:swt} shown is reproduced with permission from \citep{2018ApJ...856L..16W}.

Figure~\ref{fig:swt}, modified by \cite{2018ApJ...856L..16W}, demonstrates the application of the SWT to multi-channel solar observations, illustrating how the method enhances time-frequency resolution and enables the extraction of coherent oscillatory features across different atmospheric layers. 
The left panels display the SWT power spectra for several passbands (171~\AA, 304~\AA, H$\alpha$, 1600~\AA, 1700~\AA, and TiO), each corresponding to a distinct formation height from the chromosphere to the photosphere. 
The black dashed curves delineate the cone of influence, indicating regions where edge effects may distort the spectrum. 
Compared with CWT, the SWT performs a frequency reallocation that sharpens spectral concentration and improves mode separation, allowing clearer identification of multiple coexisting oscillation branches. 
The right panels show the reconstructed temporal components derived from the inverse SWT, which reveal the phase coherence and amplitude modulation of oscillations at different heights. 
This example highlights the strength of SWT in resolving multi-scale and multi-height oscillations within the solar atmosphere, a key capability for understanding the coupling between magnetic and acoustic processes.

The primary advantage of the SWT is its ability to enhance frequency resolution beyond that of standard wavelet analysis, producing sparse, highly localized spectral ridges that better reflect the intrinsic oscillatory modes of the signal \cite{daubechies2011synchrosqueezed}. It retains robustness against moderate noise and is particularly effective for separating closely spaced frequency components, which is crucial for disentangling overlapping solar oscillations.

However, the SWT also has limitations. The method is computationally more intensive than standard wavelet analysis, and its accuracy depends on the choice of parameters such as the mother wavelet and the discretization of frequency bins. Furthermore, unlike traditional wavelet methods, the SWT does not provide a well-established statistical significance framework comparable to the red-noise tests. Importantly, while SWT performs well for signals with relatively stationary frequency content and moderate noise levels, its performance degrades significantly in the presence of strong red noise or power-law backgrounds. The frequency reallocation procedure assumes locally sinusoidal behavior, which may not hold for stochastic processes with broad spectral content. This limitation should be carefully considered when applying SWT to solar and stellar data dominated by red-noise processes.

\section{Empirical Mode Decomposition and Hilbert--Huang Transform}
\label{sec:emd}

The Empirical Mode Decomposition (EMD) combined with the Hilbert--Huang Transform (HHT) provides a fully data-driven, adaptive approach to analyzing nonlinear and non-stationary signals. Unlike Fourier or wavelet methods, which rely on predefined bases, EMD decomposes a signal $x(t)$ into a finite set of Intrinsic Mode Functions (IMFs) derived directly from the data. Each IMF satisfies two criteria: (i) the number of extrema and zero-crossings must either be equal or differ by at most one; (ii) the mean of the upper and lower envelopes, constructed via spline interpolation of local maxima and minima, is zero. The decomposition can be expressed as
\begin{equation}
    x(t) = \sum_{i=1}^{n} c_i(t) + r_n(t),
    \label{eq:emd}
\end{equation}
where $c_i(t)$ are the IMFs and $r_n(t)$ is the final residual trend, often representing the secular background.

Once the IMFs are extracted, the Hilbert transform $\mathcal{H}[\cdot]$ is applied to each IMF $c_i(t)$ to generate an analytic signal:
\begin{equation}
    z_i(t) = c_i(t) + j \mathcal{H}[c_i(t)] = a_i(t) e^{j \phi_i(t)},
    \label{eq:analytic_signal}
\end{equation}
where $a_i(t)$ and $\phi_i(t)$ are the instantaneous amplitude and phase, respectively. The instantaneous frequency is then defined as
\begin{equation}
    \omega_i(t) = \frac{d\phi_i(t)}{dt}.
    \label{eq:instantaneous_frequency}
\end{equation}
This procedure, known as the HHT, yields a high-resolution time-frequency representation that is particularly effective for intermittent or multi-scale oscillations. Thus, EMD and HHT are closely connected: EMD provides the adaptive basis functions, and HHT extracts instantaneous frequency and amplitude information from them.

\begin{figure}[ht!]
    \centering
    \includegraphics[width=0.95\textwidth]{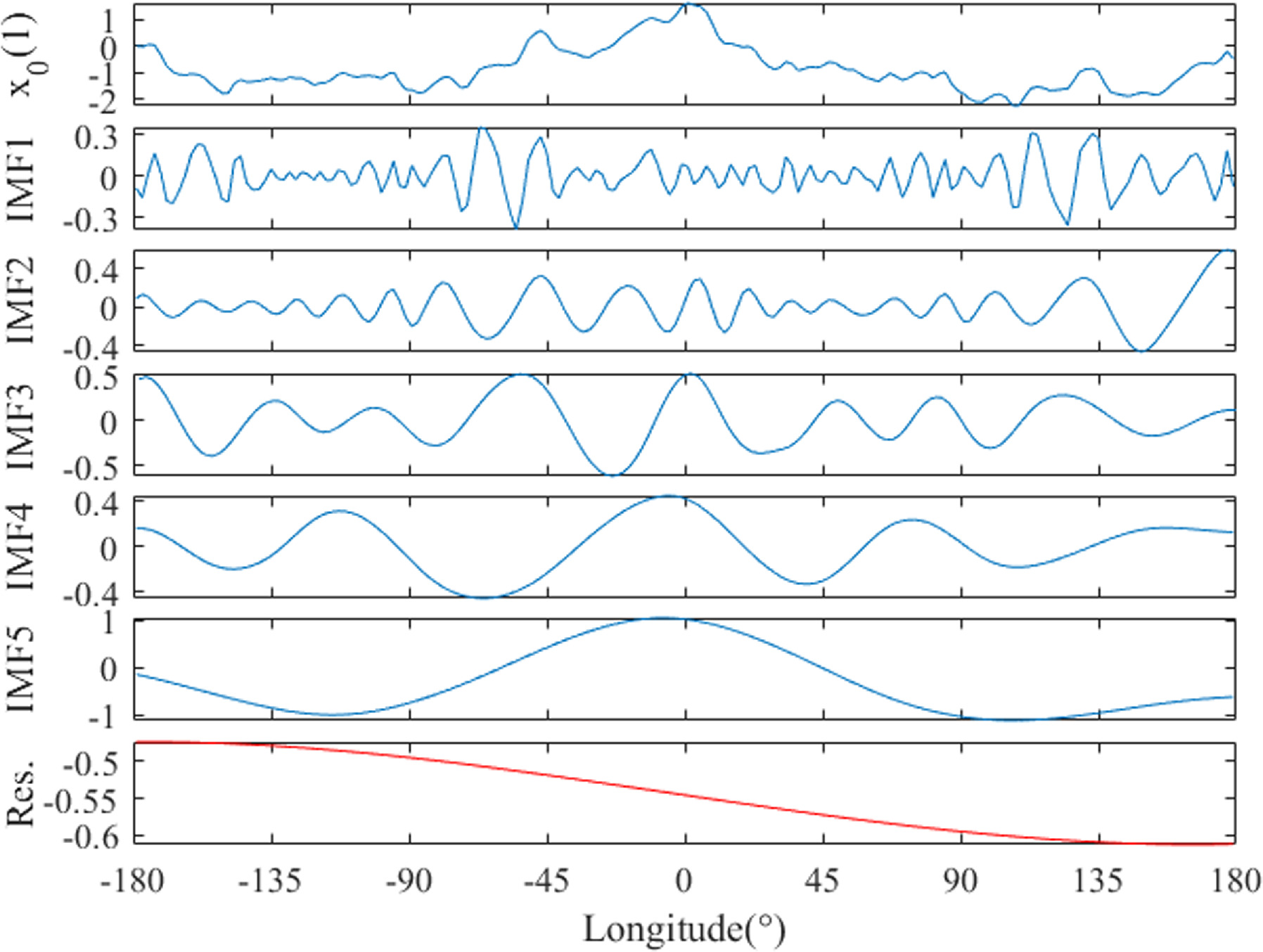}
    \caption{
        Empirical Mode Decomposition (EMD) of the mean galactic cosmic-ray signal. 
        The top panel ($x_0$) shows the original signal, while the subsequent panels display intrinsic mode functions (IMFs) from IMF1 to IMF5, representing progressively lower-frequency oscillatory components. 
        The bottom panel illustrates the final residual trend (Res.). 
        The EMD adaptively separates the signal into physically interpretable modes without assuming stationarity or linearity. 
        % Image reproduced with permission from \citep{2024ApJ...968...39C}.
        Image reproduced from \citep{2024ApJ...968...39C} under the Creative Commons Attribution 4.0 License (CC BY 4.0).
    }
    \label{fig:emd}
\end{figure}

Complete Ensemble EMD with Adaptive Noise (CEEMDAN) \cite{torres2011complete} further improves upon EEMD by introducing adaptive noise and providing a complete, unique decomposition without residual noise.
Multivariate EMD (MEMD) \cite{rehman2010multivariate} extends EMD to handle multi-channel data, which is highly relevant for solar observations involving multiple instruments or wavelengths. 
These variants significantly improve the robustness of EMD in noisy environments and enhance its interpretability in astrophysical applications.

The EMD/HHT framework has been successfully applied in various solar and stellar studies. \cite{1998RSPSA.454..903H} originally introduced the method as a powerful tool for nonlinear geophysical signals, which has since been widely adopted in astrophysics. \cite{2010PPCF...52l4009N} used EMD to investigate electromagnetic emissions generated by solar and stellar flares. \cite{2011SoPh..269..439B} demonstrated its utility in detecting variability in long-term solar activity indices. \cite{2014A&A...563A..12D} detected and identified magnetohydrodynamic sausage waves in sunspots and pores using EMD, while \cite{2016ApJ...830..110C} applied it to study quasi-periodic pulsations in the decay phase of solar and stellar flares.

Figure~\ref{fig:emd} reproduced by the reference \cite{2024ApJ...968...39C} illustrates a representative example of EMD applied to the mean galactic cosmic-ray signal, where the method decomposes the original time series into several intrinsic mode functions (IMFs) and a slowly varying residual trend. 
Each IMF corresponds to an oscillatory mode with distinct temporal scales, ranging from high-frequency fluctuations (IMF1) to large-scale variations (IMF5). 
The residual (Res.) captures the long-term monotonic trend of the dataset.     
This example highlights the key advantage of EMD in solar and astrophysical signal analysis: 
it is a fully data-driven, adaptive method that does not require a predefined basis, making it particularly effective for nonlinear and non-stationary processes. 
In the context of solar and stellar oscillations, EMD can separate different dynamical components---such as acoustic $p$-modes, magnetic oscillations, or convective fluctuations---without imposing Fourier or wavelet assumptions. 
However, as seen in practical applications, the method may suffer from mode mixing and sensitivity to noise, especially when the data contain both red and white stochastic backgrounds. 
These challenges have motivated hybrid techniques, such as ensemble EMD (EEMD) \cite{wu2009ensemble} and complementary use of synchrosqueezed transforms, to achieve more physically consistent decompositions.

Despite these successes, limitations remain. Mode mixing, sensitivity to noise, and the lack of a rigorous statistical significance framework complicate interpretation compared with Fourier or wavelet methods \cite{wu2004study}. Computational cost can also become prohibitive for very long time series. Nonetheless, with the advent of improved algorithms such as EEMD and CEEMDAN, EMD/HHT remains a uniquely powerful method for extracting intrinsic oscillatory modes from nonlinear, non-stationary solar and stellar signals.

\section{Noise Estimation and Significance Analysis}
\label{sec:noise}

Solar and stellar oscillation signals are inevitably embedded in complex noise environments that arise from both instrumental and astrophysical origins. 
Unlike idealized laboratory signals, these astrophysical time series exhibit a mixture of stochastic fluctuations, instrumental jitter, and background variations associated with turbulent convection and magnetohydrodynamic (MHD) activity. 
The resulting background noise typically combines both white noise---representing uncorrelated instrumental or photon noise---and red noise, which displays enhanced power at low frequencies due to long-memory stochastic processes \citep{2016ApJ...825..110A,2015ApJ...798....1I}. 
Such red-noise backgrounds produce power spectra that follow approximate power laws, often masking true oscillatory peaks and complicating their statistical interpretation.

To assess the reliability of detected oscillations, three principal noise estimation frameworks are commonly employed in solar and stellar physics:
(i) the False-Alarm Probability (FAP) method within the Lomb--Scargle framework, which primarily assumes white Gaussian noise;
(ii) the autoregressive (AR) approach, particularly the AR(1) model used in wavelet significance estimation; and
(iii) the Bayesian Markov Chain Monte Carlo (MCMC) framework, which models more realistic power-law or correlated noise. 
Each of these methods addresses different aspects of the noise problem, with distinct strengths and limitations, as summarized below.

\subsection{False-Alarm Probability (FAP)}
\label{sec:fap}

The Lomb--Scargle periodogram (LSP) includes a well-defined statistical framework for evaluating the significance of detected frequencies through the \textit{false-alarm probability} (FAP). 
The FAP quantifies the probability that an apparent spectral peak could arise purely from random fluctuations in white Gaussian noise. 
For a normalized LSP with $M$ independent frequencies, the probability that no peak exceeds a power level $z$ is given by:
\begin{equation}
    \mathrm{Prob}(P < z) = [1 - e^{-z}]^M,
    \label{eq:fap_prob}
\end{equation}
and hence the false-alarm probability that at least one frequency exceeds $z$ is:
\begin{equation}
    \mathrm{FAP} = 1 - [1 - e^{-z}]^M.
    \label{eq:fap}
\end{equation}
A small $\mathrm{FAP}$ (e.g., $<1\%$) indicates a statistically significant detection of a periodic component.

\begin{figure}[ht!]
    \centering
    \includegraphics[width=1\textwidth]{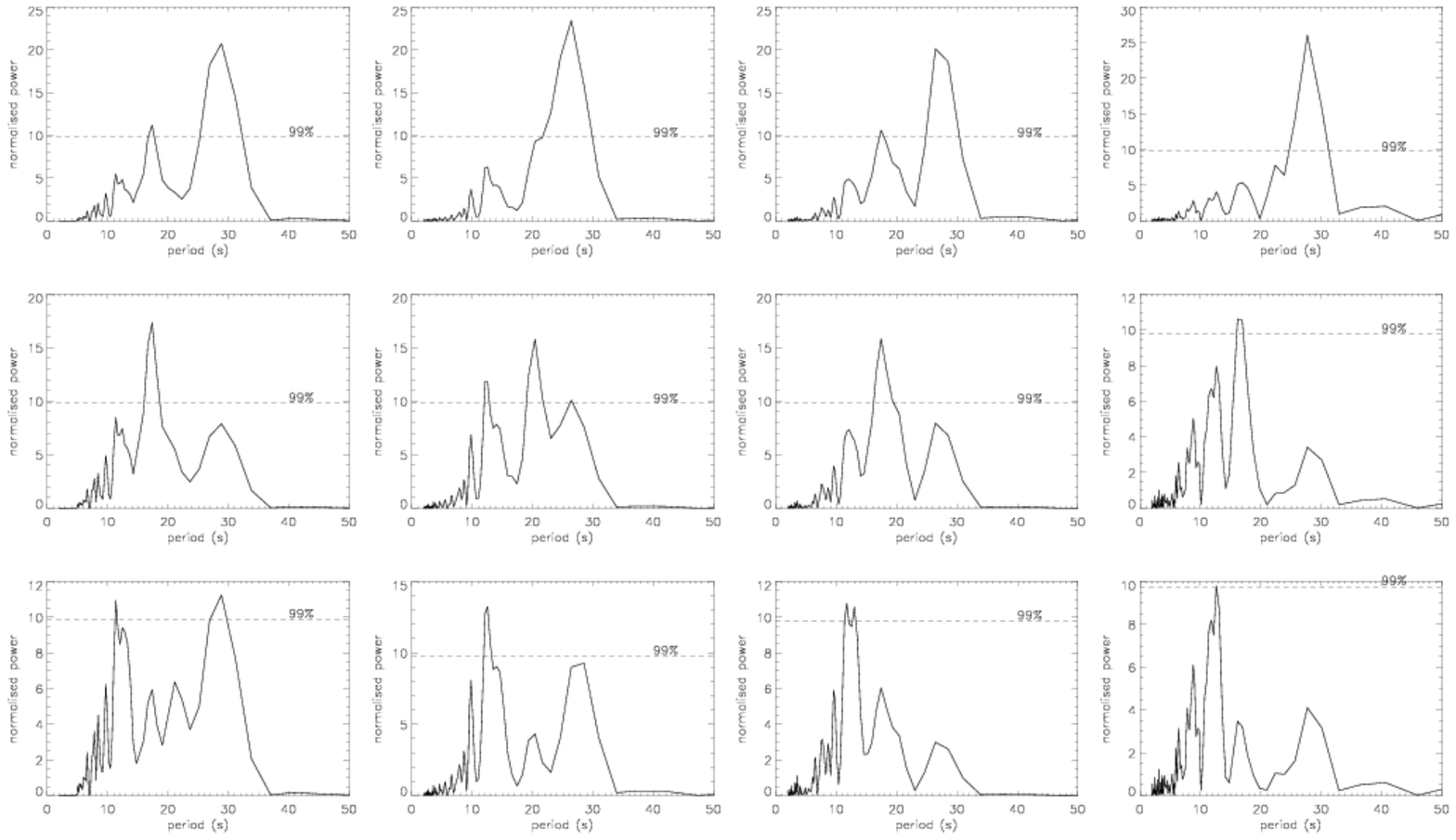}
    \caption{
        Lomb--Scargle periodograms (LSPs) illustrating the effects of iterative peak suppression at different frequencies. 
        Columns from left to right show the results for Nobeyama Radioheliograph data at 17~GHz and Nobeyama Radiopolarimeter data at 9~GHz, 17~GHz, and 35~GHz, respectively. 
        In all cases, the same three significant peaks are consistently detected above the 99\% confidence level, confirming the robustness of the detected periodicities across multiple observing frequencies. 
        % Image reproduced with permission from \citep{2009A&A...493..259I}.
        This figure is reproduced from \citep{2009A&A...493..259I}.
        Credit: Inglis \& Nakariakov, A\&A, 493, 259, 2009, reproduced with permission \textcopyright ESO.
    }
    \label{fig:ls_fap}
\end{figure}

Figure~\ref{fig:ls_fap} demonstrates the LSP application with iterative peak suppression to multi-frequency solar radio observations obtained by the Nobeyama Radioheliograph and Radiopolarimeter. 
The LSP effectively identifies periodicities even in irregularly sampled datasets and provides direct statistical evaluation through false-alarm probabilities (FAPs). 
As shown, three dominant peaks persist above the 99\% confidence level across 9, 17, and 35~GHz channels, indicating consistent oscillatory behavior over different emission heights in the solar atmosphere. 
This figure illustrates one of the major strengths of LSP analysis---its ability to extract robust, statistically validated oscillatory components without requiring evenly spaced data or interpolation. 
Such characteristics make it particularly valuable in solar radio physics, where instrumental and diurnal sampling gaps are unavoidable. 

While the FAP-based LSP is highly effective for detecting periodicities in irregularly sampled datasets, it inherently assumes a white-noise background. 
This assumption is often violated in solar observations where the noise spectrum is dominated by power-law (red) noise. 
Consequently, the LSP may overestimate the significance of low-frequency peaks, leading to false detections of quasi-periodic signals. 
Extensions of the LSP framework incorporating correlated noise covariance matrices or Bayesian formulations can partly mitigate these issues \citep{2009A&A...496..577Z,2018MNRAS.481.3083F}, but white-noise assumptions remain a key limitation.
Therefore, combining LSP with noise modeling frameworks such as AR(1) or Bayesian MCMC analysis can provide a more comprehensive assessment of signal significance in real solar observations.

\subsection{Autoregressive (AR) Noise Models in Wavelet Analysis}
\label{sec:ar1}

To analyze the time-frequency characteristics of non-stationary oscillations, wavelet analysis has become a standard tool. 
\cite{1998BAMS...79...61T} introduced the use of a first-order autoregressive process, AR(1), as a statistical background model for estimating the significance of wavelet power spectra. 
The AR(1) process is defined as
\begin{equation}
    x_t = \alpha x_{t-1} + \epsilon_t,
    \label{eq:ar1}
\end{equation}
where $\alpha$ is the autoregressive coefficient and $\epsilon_t$ represents white noise. 
This model captures the typical red-noise behavior---enhanced power at low frequencies---found in many geophysical and astrophysical signals.

In practice, the AR(1) model is fit to the observed data, and its theoretical spectrum serves as a null hypothesis to test whether the observed wavelet power at each frequency exceeds the expected noise background. 
This approach provides a convenient means of computing local (time-dependent) significance levels and has been widely adopted in solar flare and coronal oscillation analyses.

\begin{figure}[ht!]
    \centering
    \includegraphics[width=0.95\textwidth]{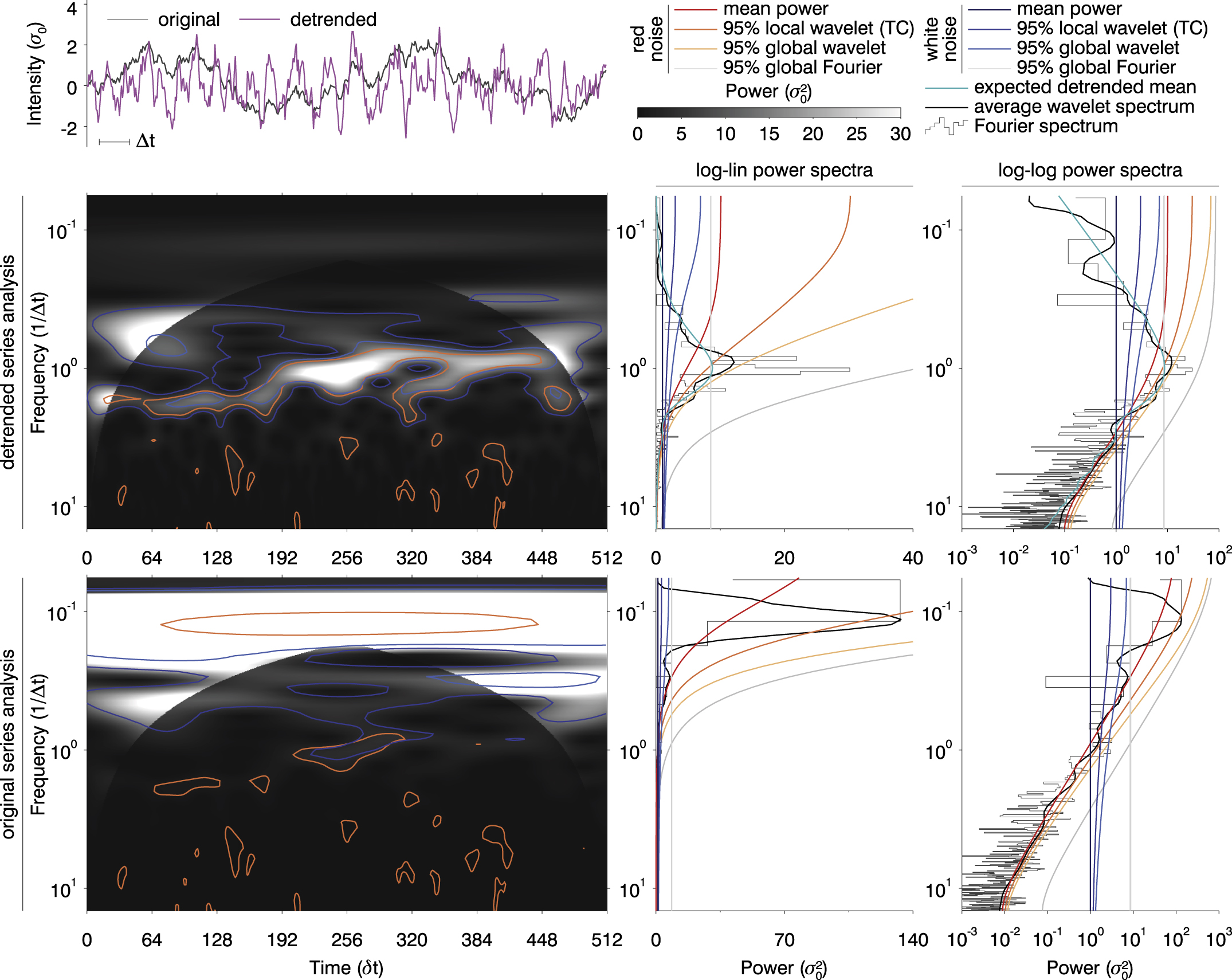}
    \caption{
        Illustration of spurious periodicities introduced by detrending a red-noise time series. 
        The top panel shows the original synthetic data (gray) that follow a $f^{-2}$ power-law spectrum, and the detrended version (magenta) obtained with a running boxcar filter of width $\delta t$. 
        The middle and bottom panels display the corresponding wavelet and Fourier power spectra, respectively, demonstrating that detrending artificially enhances high-frequency components, leading to false periodicities around the cutoff scale ($1/\delta t$). 
        The comparison of red- and white-noise reference spectra highlights how inappropriate detrending may distort the physical interpretation of quasi-periodic signals. 
        % Image reproduced with permission from \citep{2016ApJ...825..110A}.
        Image reproduced from \citep{2016ApJ...825..110A}.  \textcopyright AAS. Reproduced with permission.
    }
    \label{fig:cwt_ar}
\end{figure}

However, solar and stellar signals rarely follow a pure AR(1) process. 
Their power spectra often approximate a power law ($P(f) \propto f^{-\beta}$) rather than the Lorentzian form predicted by AR(1). 
Thus, AR(1) backgrounds systematically underestimate both low- and high-frequency power, producing biased significance levels \citep{2016ApJ...825..110A}. 
Moreover, detrending operations commonly applied to enforce AR(1)-like behavior can inadvertently introduce spurious periodicities---artificial peaks near the detrending frequency---mimicking genuine oscillations. 
An example of such artifacts is illustrated in Figure~\ref{fig:cwt_ar}.

Figure~\ref{fig:cwt_ar} illustrates a key limitation of the traditional AR(1)-based wavelet significance test when applied to detrended solar time series. 
In this example, the original synthetic signal follows a red-noise power law ($P(f) \propto f^{-2}$), yet detrending the data introduces artificial oscillations near the characteristic smoothing scale. 
As shown in both the Fourier and wavelet spectra, these spurious periodicities manifest as narrow bands of excess power at high frequencies, mimicking genuine quasi-periodic pulsations (QPPs). 
This example underscores the dual nature of detrending: while it effectively removes low-frequency trends that bias the AR(1) model, it can also generate false oscillatory power if the underlying noise follows a power-law distribution rather than a simple autoregressive process. 
Therefore, although the AR(1) background model remains convenient and computationally efficient for estimating significance in wavelet analysis, its physical validity for solar and stellar observations---where noise spectra typically exhibit scale-free behavior---is limited.

Similar conclusions have been reached in flare and QPP studies based on independent statistical frameworks. \cite{2015SoPh..290.3625S} demonstrated that impulsive soft X-ray pulsations observed during solar flares can be statistically consistent with broadband power-law variability, without requiring coherent periodic drivers. This result reinforces the notion that narrow-band enhancements may naturally arise from stochastic processes.

Furthermore, \cite{2017A&A...602A..47P} explicitly showed that by fitting a power-law red-noise model directly to the Fourier power spectrum, statistically significant QPP peaks can be recovered without applying any detrending. This provides strong evidence that detrending is not only unnecessary but potentially harmful, as it may introduce artificial periodicities near the filter cutoff scale.

\subsection{Power-law Noise and Bayesian MCMC Estimation}
\label{sec:mcmc}

To more accurately describe the stochastic nature of solar and stellar noise, modern analyses employ \textit{power-law noise models} combined with Bayesian inference. 
The power spectral density of a power-law process can be expressed as
\begin{equation}
    P(f) = A f^{-\beta} + N,
    \label{eq:power_law_noise}
\end{equation}
where $A$ is a normalization constant, $\beta$ is the spectral index, and $N$ represents the high-frequency white-noise component. 
This formulation naturally captures the scale-free, turbulent-like variability characteristic of solar coronal and photospheric dynamics.

Markov Chain Monte Carlo (MCMC) methods are used to estimate the posterior distributions of $(A, \beta, N)$ and to perform model comparisons between competing hypotheses such as AR(1), single power-law, or broken power-law models. 
\cite{2015ApJ...798..108I,2015ApJ...798....1I,2023ApJ...944...16G} demonstrated that many apparent quasi-periodic pulsations (QPPs) in flares can be explained by stochastic fluctuations superposed on a power-law background, rather than by true coherent oscillations.

\begin{figure}[ht!]
    \centering
    \includegraphics[width=1\textwidth]{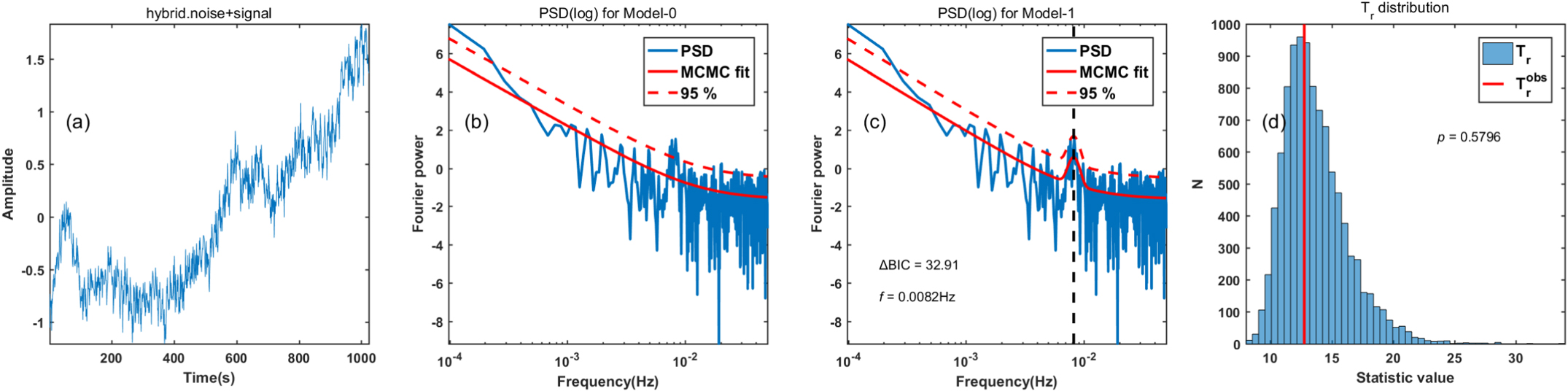}
    \caption{
        Illustration of power-law noise modeling using a Bayesian MCMC framework. 
        Panel (a) shows a synthetic signal composed of colored and white noise with a short-lived oscillation burst. 
        Panels (b) and (c) display the power spectral density (PSD) fitted by single-component ($M_0$) and double-component ($M_1$) power-law models, respectively. 
        The dashed red line marks the 95\% confidence level, while the black dashed line in panel (c) indicates the detected quasi-periodic oscillation frequency. 
        Panel (d) presents the posterior distribution of the fitted parameters, with the red vertical line showing the observed statistic value. 
        The comparison demonstrates that model $M_1$ better captures the transient oscillation on top of a power-law noise background.
        % Image reproduced with permission from \citep{2023ApJ...944...16G}.
        Image reproduced from \citep{2023ApJ...944...16G} under the Creative Commons Attribution 4.0 License (CC BY 4.0).
    }
    \label{fig:fft_mcmc}
\end{figure}

The strengths of the MCMC--Bayesian framework are its ability to:  
(i) provide full posterior uncertainty quantification for model parameters;  
(ii) rigorously test model adequacy and compare hypotheses using Bayesian evidence; and  
(iii) accommodate correlated, non-Gaussian, and scale-dependent noise structures.  
However, the approach is computationally intensive and operates in the frequency domain, thus lacking time localization. 
As shown in Figure~\ref{fig:fft_mcmc}, the method is particularly well suited for long, stationary time series where global power-law behavior dominates.

Figure~\ref{fig:fft_mcmc} illustrates the application of Bayesian MCMC methods to power-law noise modeling in solar time series. 
This approach directly fits the Fourier power spectrum using one or more power-law components, allowing rigorous estimation of the spectral slope and probabilistic assessment of candidate quasi-periodic features. 
By comparing alternative models (e.g., $M_0$ versus $M_1$), the Bayesian evidence and information criteria (such as BIC) quantify whether an observed peak reflects a genuine oscillation or simply a stochastic fluctuation within the red-noise background. 
The advantages of this approach are its statistical rigor and flexibility: it naturally handles correlated, scale-free noise and provides posterior distributions for all parameters, yielding robust confidence intervals. 
However, the MCMC framework operates entirely in the frequency domain, lacking temporal resolution. 
Consequently, it cannot identify when in time a transient oscillation occurs---a limitation that makes it complementary rather than substitutive to time-frequency approaches such as wavelet or synchrosqueezed transforms. 

In practice, Bayesian MCMC approaches are most effective when incorporated into a structured diagnostic workflow rather than applied indiscriminately. A typical analysis involves:
    (1) computing the raw power spectrum and performing an initial power-law fit to evaluate whether the background is consistent with scale-free behavior;
    (2) examining posterior predictive distributions to assess whether the fitted model reproduces the observed spectral variance;
    (3) comparing alternative models (e.g., single versus broken power-law) using Bayesian evidence or information criteria.
    Systematic residuals or frequency-dependent deviations in posterior predictive checks indicate that a simple power-law description may be inadequate. Conversely, for short or strongly non-stationary time series, MCMC-based spectral modeling can be statistically underconstrained and is best complemented by time-frequency diagnostics.

Combining MCMC-based spectral fitting with localized wavelet diagnostics offers a balanced framework for distinguishing genuine oscillatory signatures from stochastic background fluctuations in solar and stellar observations.

\subsection{Practical Guidelines and Misuse Risks}
\label{sec:practical_guidelines}

While each noise modeling approach has its mathematical foundation and application domain, practical implementation often involves subtle choices that can lead to misinterpretation or false conclusions. 
%This section outlines key guidelines and potential pitfalls for the three main noise estimation methods, addressing when AR(1) models are inappropriate, how to avoid spurious periodicities from detrending, and strategies to mitigate the computational burden of MCMC.

\subsubsection{When AR(1) Models Are Inadequate}

The AR(1) model is computationally efficient and provides closed-form significance thresholds for wavelet power. 
However, its applicability is limited to signals whose background spectrum follows a Lorentzian shape ($P(f) \propto 1/(f^2 + f_0^2)$). 
Solar and stellar noise often exhibits scale-free (power-law) behavior ($P(f) \propto f^{-\beta}$), which deviates significantly from the AR(1) expectation, especially at both low and high frequencies.

\paragraph{Identifying Inappropriate Cases:}
1. Spectral Slope Check: Compute the power spectrum of the observed data and fit a power-law index $\beta$. If $\beta$ deviates significantly from 1 (the theoretical value for AR(1)), the AR(1) model is likely inadequate.
2. Long-Memory Processes: When data exhibit long-range correlations (e.g., sunspot numbers, stellar activity cycles), the simple AR(1) model cannot capture the slowly decaying autocorrelation.
3. Multi-Scale Turbulent Backgrounds: Turbulence in the solar corona and stellar atmospheres typically produces multi-scale, non-exponential correlation structures, better described by power-law or broken power-law models.

\paragraph{Alternative Approaches:}
1. Use higher-order ARMA($p$,$q$) models, selecting the order via information criteria (AIC/BIC).
2. Adopt a power-law noise model directly, combined with Bayesian MCMC for fitting and significance testing.
3. For wavelet analysis, employ resampling (bootstrap) methods to construct empirical noise distributions, avoiding parametric assumptions about the noise form.

\subsubsection{Avoiding Spurious Periodicities from Detrending}

Detrending is commonly applied to remove low-frequency drifts before wavelet or Fourier analysis, but it can introduce artificial oscillatory signals if performed improperly. 
As shown in Figure~\ref{fig:cwt_ar}, applying a running-mean or high-pass filter to a red-noise time series creates spurious peaks near the filter's cutoff frequency.

\paragraph{Safe Detrending Principles:}
1. Avoid Over-Smoothing: Choose a smoothing window as wide as possible to minimize high-frequency artifacts. Typically, the window width should be at least twice the period of the main oscillation of interest.
2. Use Gradual Window Functions: Avoid using a rectangular (boxcar) window for smoothing; instead, use a Gaussian or cosine-bell window to reduce spectral leakage.
3. Test for Robustness: Apply multiple detrending methods (e.g., polynomial fitting, spline smoothing, empirical mode decomposition) and compare the results. An oscillation signal that consistently appears across all methods is more likely to be genuine.
4. Preserve the Original Data: Always compute reference spectra on the non-detrended data to assess distortions introduced by detrending.
5. Assess Physical Plausibility: Check if the detected periodicity is consistent with known physical processes (e.g., solar eigenmodes, Alfvén wave travel times).

\paragraph{Recommended Detrending Workflow:}
1. Visualize the raw time series to identify obvious long-term trends or drifts.
2. Remove the trend using a conservative smoothing parameter (e.g., a wide Gaussian window).
3. Compute the power spectrum of the detrended residuals and compare it with the original spectrum to check for new peaks near the cutoff frequency.
4. If necessary, employ non-parametric trend estimation methods (e.g., LOESS) to further reduce the risk of artificial periodicities.

\subsubsection{Addressing Computational Costs of MCMC Noise Modeling}

Bayesian MCMC provides rigorous uncertainty quantification and model comparison, but its computational demand can be prohibitive for long time series or real-time analysis.

\paragraph{Efficiency Bottlenecks:}
1. Parameter Space Dimensionality: Complex noise models (e.g., broken power-laws, multiple oscillatory components) increase the number of parameters, prolonging convergence time.
2. Data Length: The computational complexity of likelihood evaluation is typically $O(N\log N)$ or higher, which can be prohibitive for million-point solar observations (e.g., full-resolution SDO/AIA data).
3. Convergence Diagnostics: Running multiple chains and monitoring convergence metrics (e.g., $\hat{R}$) further increases the computational burden.

\paragraph{Acceleration Strategies:}
1. Data Compression: Apply appropriate downsampling or segment long time series, ensuring critical frequency information is preserved.
2. Efficient Samplers: Use Hamiltonian Monte Carlo (HMC) or the No-U-Turn Sampler (NUTS), which are generally more efficient than traditional random-walk MCMC in high-dimensional spaces.
3. Approximate Bayesian Computation: For particularly complex models, consider Approximate Bayesian Computation (ABC) methods that avoid direct likelihood evaluation.
4. Parallelization: MCMC chains are independent and can be run in parallel easily. Utilize multi-core CPUs or GPUs.
5. Variational Inference: Employ Variational Bayes (VB) as an approximation to MCMC. While posterior accuracy may be slightly reduced, speed can improve by 1--2 orders of magnitude.

\paragraph{Practical Recommendations:}
1. For initial exploratory analysis, use fast methods (e.g., LSP+FAP) to identify candidate frequencies, then apply MCMC for in-depth validation on shorter, selected data segments.
2. Establish standardized preprocessing and MCMC parameter configuration pipelines to ensure reproducibility.
3. Cross-validate MCMC fitting results (posterior distributions, model evidence) with simpler methods (e.g., AR(1) wavelet significance) to ensure the robustness of conclusions.

\paragraph{Future Directions:}
Integrating MCMC with time-frequency methods (e.g., wavelets, synchrosqueezed transforms) to develop time-varying Bayesian noise models represents an important pathway. This approach aims to maintain statistical rigor while capturing the evolution of non-stationary backgrounds, addressing the inherent trade-off between computational efficiency and spatio-temporal resolution.

\section{Comparative Analysis of Time-Frequency Methods}
\label{sec:comparison}

To systematically evaluate the performance of different time-frequency analysis techniques, we constructed a synthetic signal representative of solar and stellar oscillations and analyzed it using several standard and advanced spectral methods, including the FFT, CWT, EMD, and SWT. The corresponding results are shown in Figures~\ref{fig:synthetic_signal}--\ref{fig:emd_sst_reconstruction}. 

\subsection{Synthetic Signal Construction}

\begin{figure}[ht!]
    \centering
    \includegraphics[width=0.95\textwidth]{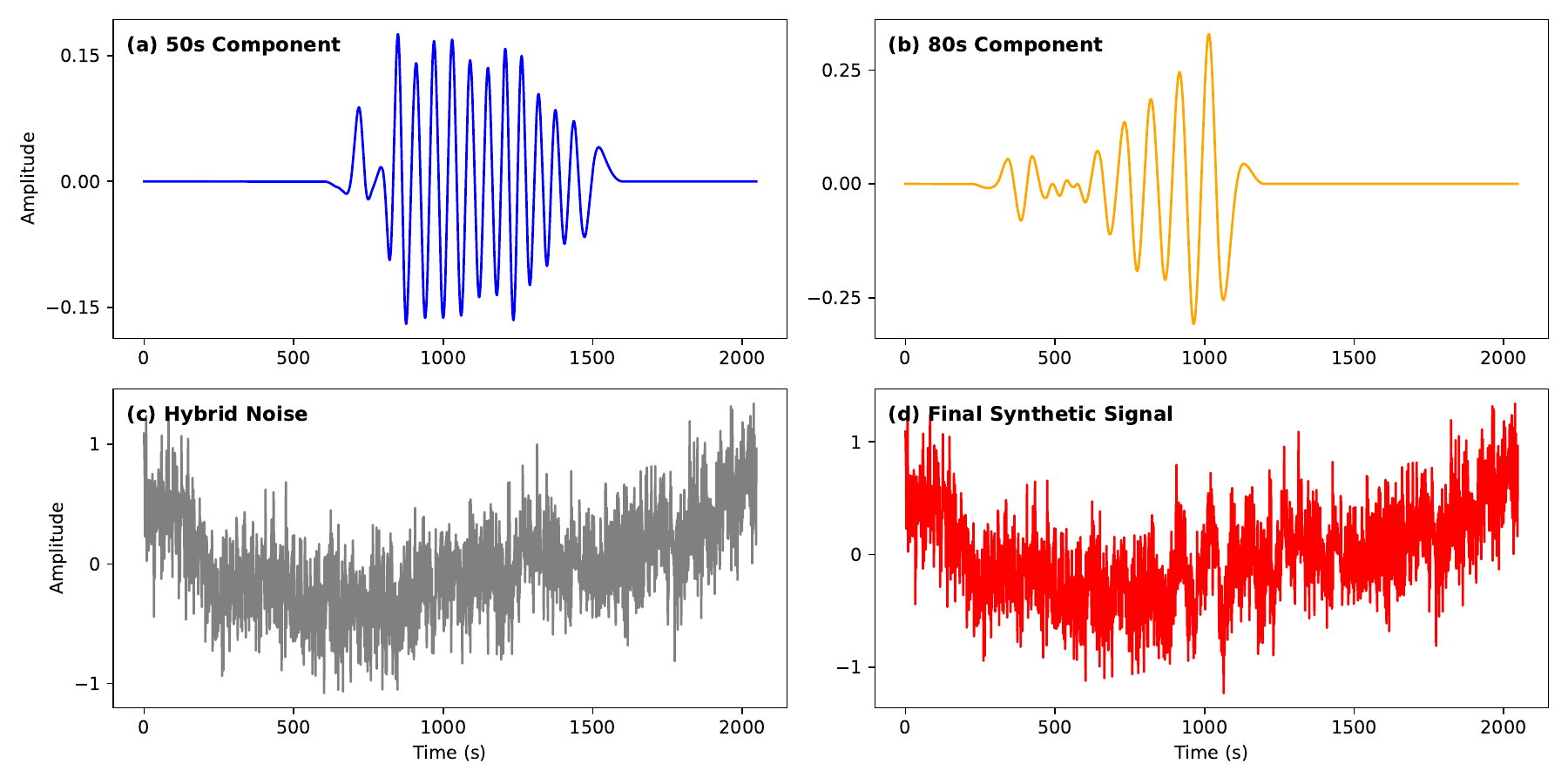}
    \caption{
        A synthetic solar-like signal: 
        (a) A 50\,s oscillation emerging between 600 and 1600\,s; 
        (b) An 80\,s oscillation emerging between 200 and 1200\,s; 
        (c) Red and white noise; 
        (d) The synthetic signal containing both oscillatory components and mixed noise.
        This synthetic signal mimics the non-stationary, multi-periodic, and noise-contaminated nature of solar and stellar oscillatory signals. It serves as a benchmark for comparing different time-frequency analysis methods.
    }
    \label{fig:synthetic_signal}
\end{figure}

The synthetic time series was designed to emulate the complex behavior of solar and stellar oscillatory signals. 
As shown in Figure~\ref{fig:synthetic_signal}, it consists of two oscillatory components with mean periods of 50\,s and 80\,s, whose amplitudes and frequencies vary in time. 
Both oscillations are superimposed on combined red and white noise components. 
The red noise follows a power-law spectrum ($P(f)\propto f^{-\beta}$, with $\beta\approx2$), representing background stochastic processes in the solar atmosphere, 
while the white noise mimics instrumental and photon noise. 

Specifically, Figure~\ref{fig:synthetic_signal}(a) shows a 50\,s oscillation emerging between 600 and 1600\,s, 
while Figure~\ref{fig:synthetic_signal}(b) displays an 80\,s oscillation appearing between 200 and 1200\,s. 
The red and white noise components are shown in Figure~\ref{fig:synthetic_signal}(c), 
and the final composite signal, combining both periodic components and noise, is presented in Figure~\ref{fig:synthetic_signal}(d). 
This signal provides a robust benchmark for testing the capability of various time-frequency methods to detect and resolve transient oscillations under noisy, nonstationary conditions. 

For completeness and reproducibility, the detailed construction procedure of the synthetic signal is described in Appendix~\ref{app:synthetic}. 
The appendix summarizes the algorithmic structure and parameter definitions implemented in the Python script used to generate the data.

\subsection{Fourier Spectral Analysis and Windowing Effects}

\begin{figure}[ht!]
    \centering
    \includegraphics[width=0.95\textwidth]{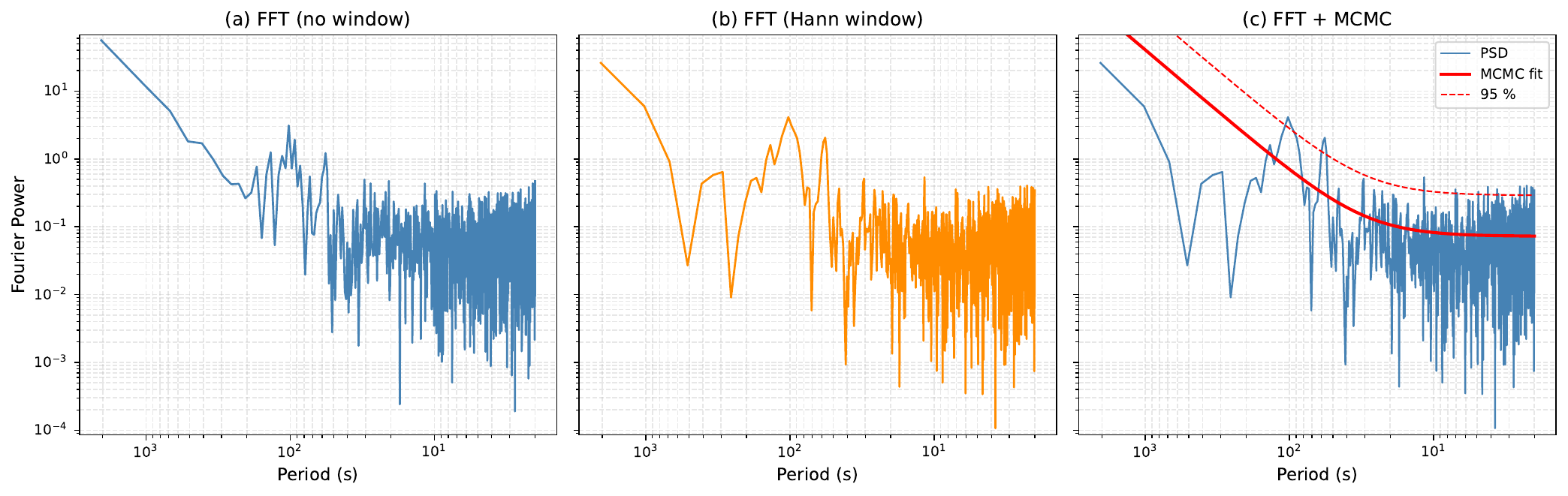}
    \caption{
        FFT spectra of the synthetic signal: 
        (a) without windowing; 
        (b) with Hann window; 
        (c) Bayesian MCMC estimation of the FFT spectrum and red-noise background \citep{2023ApJ...944...16G}.
        The FFT reveals the global frequency content of the signal. Windowing reduces spectral leakage at the cost of frequency resolution. The Bayesian MCMC approach provides a statistically rigorous estimation of the red-noise background and the significance of oscillatory peaks. For solar and stellar data, FFT is suitable for stationary oscillations but fails to capture transient events. The Bayesian MCMC method improves noise modeling but is computationally intensive and lacks temporal localization.
    }
    \label{fig:windows_mcmc}
\end{figure}

Figure~\ref{fig:windows_mcmc} presents the Fourier power spectra of the synthetic solar-like signal obtained using three approaches: (a) without windowing, (b) with a Hanning window, and (c) with Bayesian noise estimation based on MCMC sampling. 
These three panels illustrate the essential trade-offs involved in practical frequency-domain analysis of solar and stellar oscillations.

The first case, shown in Figure~\ref{fig:windows_mcmc}(a), corresponds to the raw DFT using a rectangular window. 
Its main advantage is the preservation of the true frequency resolution: because no additional weighting is applied, the spectral peaks remain sharp, making it possible to distinguish closely spaced frequencies or narrowband oscillations. 
However, the discontinuities at the edges of the finite observation window produce spectral leakage, spreading the energy of strong frequencies into adjacent bins. 
This leakage may generate false low-amplitude peaks and obscure weak oscillatory components, especially when the signal contains strong transient or noise-contaminated intervals.

In Figure~\ref{fig:windows_mcmc}(b), a Hanning window is applied prior to the Fourier transform to reduce the effect of spectral leakage. 
The Hanning window smoothly tapers the signal toward zero at both ends, effectively minimizing the edge discontinuities and confining the main lobe energy. 
As a result, the side-lobe amplitudes are substantially suppressed, improving the reliability of peak identification in noisy spectra. 
However, the trade-off is a loss of frequency resolution---the main lobes of the spectral peaks broaden, causing nearby modes to merge or appear less distinct. 
This trade-off between leakage suppression and resolution is a general characteristic of windowing in Fourier analysis, where rectangular windows provide maximum resolution but poor leakage control, and tapered windows (e.g., Hanning, Hamming) provide smoother spectra at the expense of precision.

Figure~\ref{fig:windows_mcmc}(c) shows the result of a Bayesian power-spectrum estimation using the MCMC method. 
In this approach, the background noise is modeled as a power-law function of frequency,
\begin{equation}
    P(f) = A f^{-\beta} + N,
    \label{eq:bayesian_psd}
\end{equation}
and the parameters $(A, \beta, N)$ are estimated by sampling their posterior probability distributions. 
This statistical framework accounts for both red and white noise components simultaneously, providing a more physically motivated background model than traditional ad hoc smoothing. 
The key advantages of the MCMC-Bayesian approach are its ability to (i) rigorously quantify parameter uncertainties, (ii) evaluate the statistical significance of spectral peaks, and (iii) discriminate between genuine oscillatory power and noise-driven fluctuations. 

Nevertheless, this approach also has limitations. 
First, it assumes a stationary background model and therefore lacks temporal localization; transient events cannot be isolated in time as they can with time-frequency methods such as wavelets. 
Second, MCMC inference is computationally expensive and may require careful tuning of priors and convergence criteria. 
Finally, the accuracy of the noise model depends on the validity of the assumed power-law form; departures from this assumption can lead to biased spectral fits. 
Despite these caveats, Bayesian MCMC-based spectral estimation represents one of the most robust and interpretable approaches for quantifying oscillatory power in noisy solar data.

\subsection{Continuous Wavelet Transform and Detrending Effects}

\begin{figure}[ht!]
    \centering
    \includegraphics[width=0.95\textwidth]{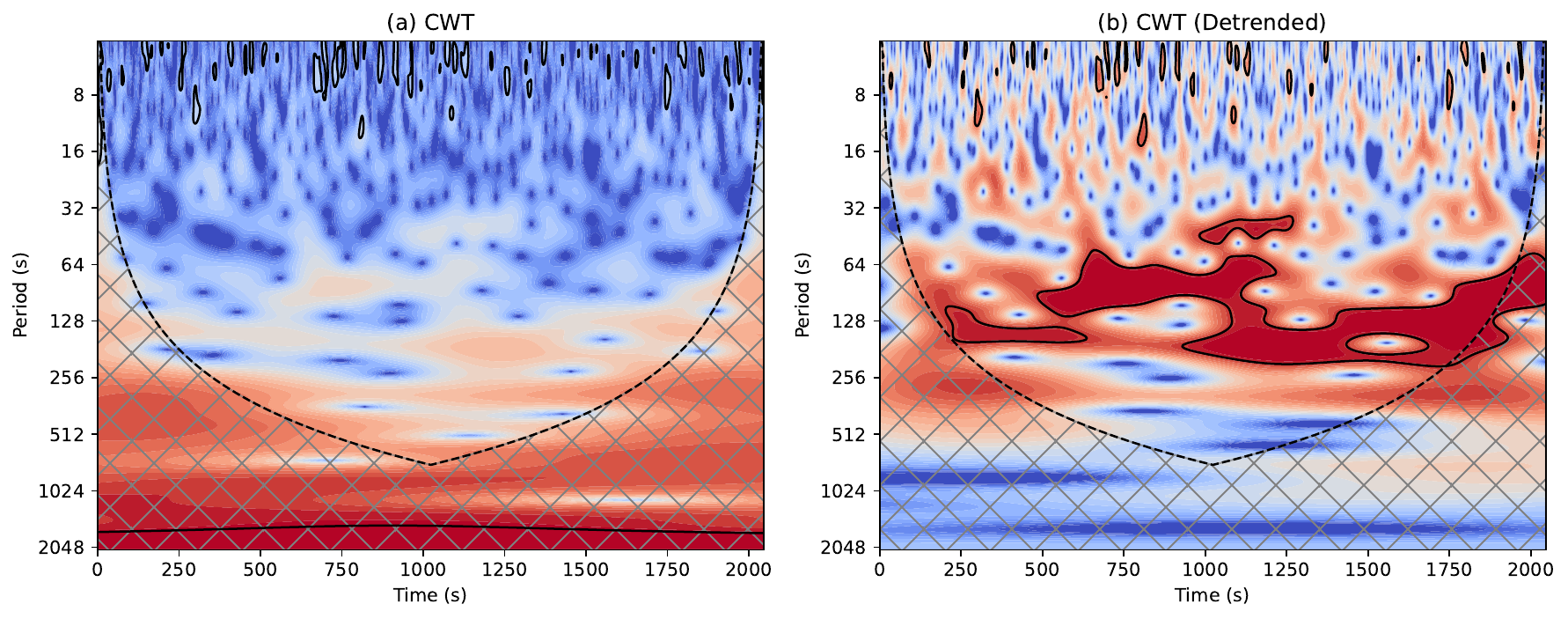}
    \caption{
        Continuous wavelet transform (CWT) analysis of the synthetic signal: 
        (a) CWT spectrum without detrending; 
        (b) CWT significance levels estimated with an AR(1) red-noise background following \cite{1998BAMS...79...61T}.
        The CWT reveals the time-frequency evolution of oscillatory power. Without detrending, the red-noise background dominates the low-frequency power, masking the oscillations. Detrending enhances the visibility of oscillations but may introduce spurious periodicities. For solar and stellar data, CWT is effective for non-stationary and transient oscillations, but the detrending process must be carefully chosen to avoid artifacts.
    }
    \label{fig:cwt_detrend_undetrend}
\end{figure}

Figure~\ref{fig:cwt_detrend_undetrend} compares the continuous wavelet transform (CWT) spectra of the synthetic solar-like signal before and after detrending. 
Panel~(a) shows the original signal analyzed without detrending, while panel~(b) presents the result after removing the slowly varying background using temporal smoothing and subtraction. 
Both analyses adopt the Torrence and Compo (1998) framework with an AR(1) red-noise background model.

In the undetrended case (Figure~\ref{fig:cwt_detrend_undetrend}a), the CWT spectrum is dominated by large-scale, low-frequency power that originates from the red-noise background rather than from coherent oscillatory modes. 
Because the AR(1) background model assumes a first-order autoregressive stochastic process, it cannot completely remove the power-law-like noise present in the signal. 
Consequently, the oscillatory components embedded within the signal---particularly the 50~s and 80~s periodicities---are partially masked by the broad continuum of red noise. 
This leads to an underestimation of the significance of genuine oscillations, especially when their amplitudes are small compared with the background fluctuations. 
Therefore, without detrending, the CWT tends to overestimate low-frequency power while failing to isolate short-lived or moderate-amplitude oscillatory structures.

In contrast, the detrended result (Figure~\ref{fig:cwt_detrend_undetrend}b) shows that the 50~s and 80~s oscillatory components become much more distinct. 
By applying a smoothing filter to remove the long-term background trend, the non-oscillatory variations that dominate the red-noise continuum are effectively suppressed. 
This enhances the relative contrast of the true periodic components and allows them to exceed the 95\% confidence level in the AR(1) significance test. 
However, detrending also introduces side effects: the subtraction of a smoothed background may distort the low-frequency components and produce artificial periodicities. 
In the present example, spurious power appears near periods of $\sim128$~s and $\sim256$~s, which do not correspond to any actual oscillations in the synthetic signal. 
Such artifacts arise because detrending acts as a high-pass filter, altering the signal spectrum and generating residual oscillations at scales comparable to the cutoff frequency of the smoothing function.

These results highlight an intrinsic trade-off in CWT-based analysis. 
Undetrended signals preserve the full spectral content but suffer from contamination by long-term trends and red noise, whereas detrending improves the detectability of localized oscillations at the expense of introducing potential artifacts. 
For real solar and stellar observations---where both background evolution and stochastic noise are present---the choice of whether and how to detrend must therefore be made carefully, ideally by combining CWT analysis with independent noise modeling or statistical tests to ensure the physical authenticity of detected oscillations.

\subsection{Adaptive Decomposition: EMD and SWT}
\begin{figure}[ht!]
    \centering
    \includegraphics[width=0.95\textwidth]{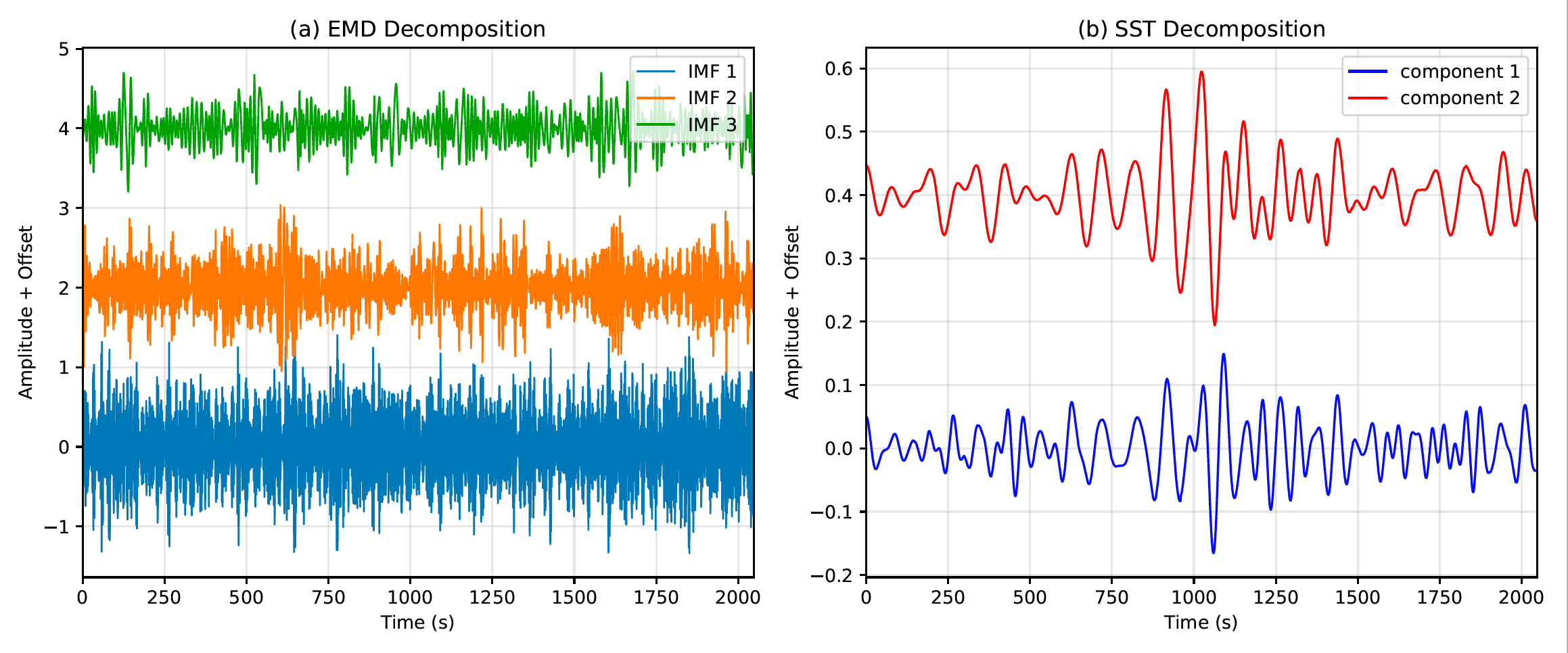}
    \caption{
        Adaptive decomposition of the synthetic signal:
        (a) HHT based on EMD showing three intrinsic mode functions (IMFs);
        (b) SWT decomposition with two reconstructed oscillatory components showing higher spectral concentration.
        EMD adaptively decomposes the signal into oscillatory modes without a predefined basis, but it is sensitive to noise and may suffer from mode mixing. SWT sharpens the time-frequency representation by reassigning energy, providing better separation of multi-component oscillations. For solar and stellar data, these methods are useful for nonlinear and non-stationary signals, but they lack rigorous statistical significance tests and may be affected by power-law noise.
    }
    \label{fig:emd_sst_reconstruction}
\end{figure}

Adaptive data-driven techniques such as the Hilbert--Huang Transform (HHT) based on Empirical Mode Decomposition (EMD) and the Synchrosqueezed Wavelet Transform (SWT) provide flexible frameworks for analyzing nonlinear and nonstationary oscillations. 
Unlike Fourier or wavelet methods, these approaches adaptively extract oscillatory modes from the data itself without assuming a predefined basis, offering valuable insight into time-varying frequency behavior.

As illustrated in Figure~\ref{fig:emd_sst_reconstruction}(a), the EMD decomposes the synthetic signal into three intrinsic mode functions (IMFs). 
The first two IMFs capture the 50\,s and 80\,s oscillatory components embedded in the signal, while the third represents the background trend. 
Although EMD effectively identifies major oscillatory scales, it is sensitive to the presence of power-law noise (red noise), which causes \textit{mode mixing}---the blending of physical and noise-driven oscillations across IMFs. 
In this case, the red and white noise components induce artificial amplitude modulation and frequency drift, reducing the accuracy of true oscillatory component separation. 
Furthermore, EMD lacks a rigorous statistical framework for noise significance testing, making it difficult to distinguish physical oscillations from stochastic fluctuations.

The SWT shown in Figure~\ref{fig:emd_sst_reconstruction}(b) enhances the time-frequency concentration by reallocating wavelet coefficients through synchrosqueezing. 
This refinement sharpens spectral ridges and allows for partial reconstruction of oscillatory components. 
However, in the presence of a steep power-law background, the SWT may also overemphasize high-frequency noise or split broad oscillations into multiple narrow ridges, leading to interpretation ambiguities. 
Additionally, the decomposition quality depends on manually selected frequency bands, introducing subjectivity into the analysis.

Overall, both EMD and SWT offer strong advantages for analyzing transient or nonlinear oscillations compared to stationary methods. 
However, neither approach can fully eliminate or model the effects of power-law noise typical of solar and stellar signals. 
EMD provides direct time-domain decomposition but is noise-sensitive and nonunique, whereas SWT yields sharper spectral localization but requires careful parameter selection and prior knowledge of frequency ranges. 
Future developments combining adaptive decomposition with Bayesian or statistical noise modeling may provide more physically meaningful separation of true oscillatory modes from stochastic solar background variability.

\subsection{Summary of Comparative Analysis}

Figures~\ref{fig:synthetic_signal}--\ref{fig:emd_sst_reconstruction} collectively demonstrate the comparative performance of various spectral and time-frequency analysis techniques when applied to synthetic solar-like signals embedded in hybrid white and red noise. 
The results highlight that each method possesses unique advantages and limitations depending on the degree of stationarity, noise characteristics, and temporal variability of the data.

The classical FFT provides global frequency information with high precision but assumes stationarity and cannot capture transient or evolving oscillations. 
The Bayesian FFT+MCMC approach improves statistical rigor and noise modeling by fitting power-law spectra, yet it sacrifices temporal localization. 
The CWT resolves time-varying oscillations and localizes power in both time and frequency, but it is sensitive to detrending and edge effects (the cone of influence). 
Adaptive methods such as EMD and HHT decompose nonlinear signals into intrinsic mode functions without requiring a basis, but they are prone to mode mixing and perform poorly in the presence of power-law noise. 
The SWT refines wavelet analysis by reassigning time-frequency energy, offering improved sharpness and partial reconstruction capability, but its performance degrades when the background noise spectrum deviates strongly from white-noise assumptions.

These comparisons demonstrate that no single method offers a complete solution for solar and stellar oscillations, which are inherently nonstationary, multi-periodic, and embedded in correlated (red) noise. 
The optimal choice of technique must therefore be guided by the scientific objective and the statistical nature of the dataset.

\section{Conclusions and Outlook}
\label{sec:discussion}
The comparative evaluation of time-series and spectral analysis methods presented in this review highlights the diverse methodological landscape available for studying solar and stellar oscillatory signals. 
A central theme that emerges is the inherent trade-off between temporal localization and frequency precision, governed by the uncertainty principle and further modulated by the specific assumptions of each technique. 
This trade-off dictates that the choice of analytical method is not one of absolute superiority, but rather of appropriate application, guided by the specific scientific question, the characteristics of the data, and the nature of the underlying physical processes.

Our analysis underscores that no single method provides a universal solution. 
Classical Fourier-based methods, including FFT and Lomb-Scargle periodograms, remain indispensable for their statistical rigor and efficiency in extracting global frequency content from stationary or quasi-stationary signals. 
However, their reliance on stationarity and uniform sampling limits their utility for the transient, intermittent, and non-stationary oscillations prevalent in solar flares, coronal waves, and stellar outbursts. 
For these phenomena, time-frequency methods such as the Continuous Wavelet Transform (CWT) and the Synchrosqueezed Wavelet Transform (SWT) offer critical insights by revealing the temporal evolution of spectral power. 
Adaptive methods, EMD/HHT, further extend this capability to nonlinear and non-stationary signals without predefined bases, albeit at the cost of potential mode mixing and a less formal statistical framework.

A paramount and persistent challenge in this field is the accurate characterization and mitigation of background noise. 
Solar and stellar signals are embedded in complex, often scale-free (red or power-law) stochastic backgrounds generated by turbulent convection, magnetohydrodynamic processes, and instrumental effects. 
As demonstrated, improper handling of this noise—such as inappropriate detrending or applying white-noise significance tests to red-noise data—can lead to two opposing but equally problematic outcomes: the generation of spurious periodicities or the suppression of genuine weak oscillations. 
Therefore, moving beyond simple white-noise assumptions is not merely an academic refinement but a necessity for reliable astrophysical inference.

Looking ahead, the future of solar and stellar oscillation analysis lies in the strategic integration and advancement of existing frameworks, driven by both observational needs and computational innovation. 
We identify several key directions:

\begin{enumerate}
    \item \textbf{Integrated Noise Modeling:} Future analysis pipelines must move away from treating noise removal as a separate preprocessing step. Instead, physically motivated noise models (e.g., power-law, autoregressive, or Harvey-like profiles) should be embedded directly into the signal decomposition or detection algorithms. Bayesian frameworks are particularly promising here, as they allow for simultaneous parameter estimation and model comparison within a unified probabilistic context.
    
    \item \textbf{Hybrid Time-Frequency-Bayesian Methods:} Bridging the gap between the temporal localization of wavelets and the statistical rigor of Bayesian inference represents a major frontier. Developing formal significance tests for CWT, SWT, and HHT within a Bayesian or frequentist framework that accounts for correlated noise would dramatically enhance the reliability of these methods. Similarly, extending Bayesian MCMC techniques to provide time-resolved (e.g., wavelet-MCMC) analyses could offer the best of both worlds: robust uncertainty quantification with temporal evolution.
    
    \item \textbf{Computational Efficiency and Reproducibility:} As data volumes from missions like \textit{Solar Orbiter}, DKIST, and PLATO explode, computational efficiency becomes critical. This necessitates the development of optimized algorithms, potential hardware acceleration (e.g., GPU-based FFT/wavelets), and approximate Bayesian computation techniques. Furthermore, the establishment of standardized, open-source analysis pipelines and benchmark datasets will be crucial for ensuring reproducibility, facilitating comparative studies, and accelerating community-wide methodological progress.
    
    \item \textbf{Physics-Informed Machine Learning (PIML):} The application of deep learning for automated feature detection (e.g., QPPs, oscillation modes) holds great promise. The future, however, lies in \textit{Physics-Informed Machine Learning}, where the interpretable features extracted by classical methods (wavelet coefficients, IMFs, periodogram peaks) are used to inform and constrain neural network models. This synergy can combine the pattern recognition power of AI with the physical insight of traditional signal processing, moving beyond ``black-box'' predictions toward physically interpretable models.
    
    \item \textbf{Multi-Dimensional and Multi-Messenger Analysis:} The analysis must evolve from 1D time series to multi-dimensional data cubes (space, time, wavelength) to fully exploit modern observations. Techniques for analyzing propagating waves in imaging data, coupling different atmospheric diagnostics, and integrating multi-messenger information (e.g., light curves with spectral line profiles or magnetic field vectors) are essential for a holistic understanding of energy transport and wave coupling in stellar atmospheres.
\end{enumerate}

In summary, the path forward is not to seek a single superior method, but to cultivate a synergistic methodological ecosystem. 
The combined strengths of Fourier analysis, wavelet and synchrosqueezed transforms, adaptive decomposition, and Bayesian statistics, when carefully selected and potentially hybridized, offer a powerful toolkit. 
This integrated approach will be fundamental for transforming high-cadence, multi-wavelength observations into profound insights about magnetized plasma dynamics, energy release processes, and stellar structure. 
By advancing both our analytical techniques and our physical noise models, we can move closer to achieving the ultimate goal: a precise, reliable, and physically grounded understanding of the complex oscillatory symphony that pervades the Sun and stars.

% ============================================================
\appendix
\section{Construction of the Synthetic Signal}
\label{app:synthetic}
% ============================================================

The synthetic signal used throughout this study was generated using a dedicated Python script. 
Rather than providing the full source code, we summarize here the main design principles and algorithmic steps to ensure clarity and reproducibility.

The signal is constructed as a linear superposition of three components: (1) Two time-localized oscillatory components with prescribed mean periods (50\,s and 80\,s).  Each oscillation is modeled as a sinusoidal function whose amplitude and instantaneous frequency vary smoothly in time.  Temporal localization is achieved through envelope functions that restrict the oscillations to specific time intervals.
 (2) A red-noise background generated to follow a power-law power spectrum 
    ($P(f)\propto f^{-\beta}$ with $\beta\approx2$), 
    mimicking stochastic processes commonly observed in solar atmospheric data.
 (3) An additive white-noise component drawn from a Gaussian distribution 
    to represent instrumental and photon noise.

All parameters, including sampling cadence, total duration, oscillation periods, 
time intervals of activity, noise amplitudes, and spectral index,  are explicitly defined within the script to allow controlled experiments. 
The final synthetic time series is obtained by summing the oscillatory components and noise terms,  providing a well-defined benchmark with known ground-truth temporal and spectral properties.

\vskip8pt

\ack{
DY was supported by the National Natural Science Foundation of China (NSFC,12173012, 12473050), the Guangdong Natural Science Funds for Distinguished Young Scholars (2023B1515020049), the Shenzhen Science and Technology Project (JCYJ20240813104805008) the Shenzhen Key Laboratory Launching Project (No. ZDSYS20210702140800001).
SF is supported by the Open Fund of the Yunnan Key Laboratory of Computer Technologies Application.
}

\bibliographystyle{RS} %%%% .BST file
\bibliography{2025-feng_v1} %%%%% .Bib file

\end{document}